\documentclass[twocolumn,english,prb]{revtex4}
\usepackage[T1]{fontenc}
\usepackage[latin9]{inputenc}
\usepackage{babel}
\usepackage{amstext}
\usepackage{amssymb}
\usepackage{cancel}
\usepackage{graphicx}
\usepackage{esint}
\PassOptionsToPackage{normalem}{ulem}
\usepackage{ulem}
\usepackage[unicode=true,
 bookmarks=true,bookmarksnumbered=false,bookmarksopen=false,
 breaklinks=false,pdfborder={0 0 1},backref=false,colorlinks=false]
 {hyperref}
\hypersetup{pdftitle={Hierarchy of correlations},
 pdfauthor={Álvaro Gómez-León},
 pdfsubject={manuscript}}
\usepackage{breakurl}

\makeatletter
\@ifundefined{textcolor}{}
{%
 \definecolor{BLACK}{gray}{0}
 \definecolor{WHITE}{gray}{1}
 \definecolor{RED}{rgb}{1,0,0}
 \definecolor{GREEN}{rgb}{0,1,0}
 \definecolor{BLUE}{rgb}{0,0,1}
 \definecolor{CYAN}{cmyk}{1,0,0,0}
 \definecolor{MAGENTA}{cmyk}{0,1,0,0}
 \definecolor{YELLOW}{cmyk}{0,0,1,0}
}

\makeatother

\begin{document}

\title{Hierarchy of correlations: Application to Green's functions and interacting
topological phases}

\author{Álvaro Gómez-León}

\affiliation{Department of Physics and Astronomy and Pacific Institute of Theoretical
Physics\\
University of British Columbia, 6224 Agricultural Rd., Vancouver,
B.C., V6T 1Z1, Canada}
\email{agomez@phas.ubc.ca}

\date{\today}
\begin{abstract}
We study the many-body physics of different quantum systems using
a hierarchy of correlations, which corresponds to a generalization
of the $1/\mathcal{Z}$ hierarchy. The decoupling scheme obtained
from this hierarchy is adapted to calculate double-time Green's functions,
and due to its non-perturbative nature, we describe quantum phase
transition and topological features characteristic of strongly correlated
phases. As concrete examples we consider spinless fermions in a dimerized
chain and in a honeycomb lattice. We present analytical results which
are valid for any dimension and can be generalized to different types
of interactions (e.g., long range interactions), which allows us to
shed light on the effect of quantum correlations in a very systematic
way. Furthermore, we show that this approach provides an efficient
framework for the calculation of topological invariants in interacting
systems.
\end{abstract}
\maketitle
\tableofcontents{}

\section{Introduction}

The study of quantum many-body systems is a very active research field,
mainly due to the large number of fascinating effects that can be
observed. The main reason for the large degree of complexity in these
systems is the presence of non-local correlations between particles.
The development of different methods to extract the relevant information
for their description is then crucial, and an important source of
progress and debate in the field. While many standard techniques such
as diagramatic, bosonization\cite{Bosonization}, dynamical mean field
theory\cite{DynamicalMFT1,DynamicalMFT2}, etc, have been applied,
it is not always clear how correlations are treated in these approximations,
and this can cause problems in the description of the system.

A very important feature of quantum many-body systems is the existence
of quantum phase transitions (QPT)\cite{QPT-Sachdev}. They describe
a change in the many-body ground state properties of a quantum system
when a physical parameter is tuned. Unlike classical phase transitions,
which are driven by thermal fluctuations, QPT are driven by quantum
fluctuations, and are strictly speaking, $T=0$ phase transitions.
The characterization of phase transitions has notably evolved, starting
from the description of classical phase transitions by Ehrenfest,
to the characterization and classification of order parameters by
Landau\cite{PhaseTransitions-Landau}, or the modern dynamical mean
field theories (not to mention the crucial renormalization group theory
developed by Wilson\cite{Renormalization-Wilson}). Crucially, it
has been shown that the non-analyticity of the transitions is ubiquitous
in all these approaches, meaning that perturbative methods (in the
sense of standard perturbation theory) are not generally enough for
their description.

Another important feature of many-body systems, intensively studied
in the last decade, is topological phase transitions(TPT)\cite{QPT-Bernevig,RevModPhys-TPT}.
In general, the topological phases cannot be characterized using an
order parameter, which makes them different from the usual QPT, and
one needs to calculate topological invariants. Furthermore, their
bulk spectral properties cannot be differentiated from those of trivial
phases, and only when looking at the boundary of the system, can one
observe the physical differences due to the presence of edge modes.
Although all the main properties and possible topological phases are
well understood for the case of non-interacting systems\cite{Ludwig-Classification},
the effect of interactions is still unclear and requires further studies.
Nevertheless, huge progress has been made in the last years, and numerous
phases with very interesting properties have been predicted\cite{Interac-TI-Class,Top-Mott-Ins,Top-Kondo-Ins,Frac-QHI1,Frac-QHI2,Floquet-FCI,Xin-Li2015}.

A very interesting method used to characterize magnetic systems -archetypal
examples in the theory of quantum phase transition\cite{TransverseIsing}-
is the $1/\mathcal{Z}$ expansion, where $\mathcal{Z}$ is the coordination
number of the system\cite{Z-Exp1,Z-Exp2}. In this case, one assumes
that contributions from the interaction part of the Hamiltonian scale
inversely proportional to $\mathcal{Z}$, providing a hierarchical
expansion of the different statistical averages. From this very general
argument, it is clear that the hierarchy will be accurate in high
dimensional systems (where the coordination number is large), which
is the reason why it has been mainly applied in the description of
3D materials. Studies of electronic and bosonic systems with itinerant
particles have also been recently analyzed using this approach \cite{Queisser}.

In this work we discuss the origin of the coordination number scaling,
and show that it can be derived from the monogamy property of entanglement.
This feature makes the $1/\mathcal{Z}$ expansion equivalent to an
expansion in terms of spatial correlations, and their scaling allows
to approximate the $n$-point functions. This connection allows us
to generalize the hierarchy to $\mathbf{k}$ space, and obtain a different
scaling for $\mathbf{k}$ space correlations. We show that for low
dimensional systems it is more effective to study $\mathbf{k}$ space
correlations, as they provide a faster convergence for the hierarchy.
We prove explicitly these results in a honeycomb lattice\cite{PhaseDiag-HC3,PhaseDiag-HC2,PhaseDiag-HC1,ExactDiag-HC,HC-Interactions1}
and in a dimerized 1D chain, both populated by spinless fermions.
Finally, we take advantage of the non-perturbative nature of this
approach to analyze the topological properties of the dimerized chain.
We show that the hierarchy of correlations allows to extract the topological
invariants beyond the non-interacting limit.

\section{Hierarchy of correlations}

The conventional assumption of the $1/\mathcal{Z}$ hierarchy is that
the interaction part of the Hamiltonian, proportional to $V$, scales
as $1/\mathcal{Z}$, where $\mathcal{Z}$ is the coordination number
of the system. At first glance, this can be interpreted as an expansion
in the interaction part of the Hamiltonian; however this is not the
case, and the expansion is not done in terms of a small parameter,
but rather in terms of decreasing contributions of correlations to
the density matrix. Formally, we can derive the hierarchy as follows:
Let us introduce the density matrix of a system $\rho$, and the reduced
density matrix $\rho_{\mathbf{x}}=\textrm{Tr}_{\bcancel{\mathbf{x}}}\left\{ \rho\right\} $,
where $\textrm{Tr}_{\bcancel{\mathbf{x}}}\left\{ \rho\right\} $ means
trace over all degrees of freedom but $\mathbf{x}$ ($\mathbf{x}$
corresponds to a specific lattice site for the case of the $1/\mathcal{Z}$
hierarchy). If we consider the 2-point reduced density matrix $\rho_{\mathbf{x},\mathbf{y}}=\textrm{Tr}_{\bcancel{\mathbf{x}},\bcancel{\mathbf{y}}}\left\{ \rho\right\} $,
we can rewrite this quantity as follows:
\begin{equation}
\rho_{\mathbf{x},\mathbf{y}}=\rho_{\mathbf{x}}\rho_{\mathbf{y}}+\rho_{\mathbf{x},\mathbf{y}}^{C}\label{eq:Decomposition}
\end{equation}
where $\rho_{\mathbf{x},\mathbf{y}}^{C}$ is just the difference between
the 2-point density matrix, and the product of 1-point density matrices
at each point $\rho_{\mathbf{x}}\rho_{\mathbf{y}}$ (i.e., $\rho_{\mathbf{x},\mathbf{y}}^{C}$
contains the correlations between the two sites). If we assume the
scaling $\rho_{\mathbf{x},\mathbf{y}}^{C}\sim1/\mathcal{Z}$ (this
is motivated in terms of the entanglement monogamy in the next paragraph),
we find that the different statistical averages can be separated into
their uncorrelated part and contributions due to correlations, which
decay with different inverse powers of the coordination number $\mathcal{Z}$.
This defines the $1/\mathcal{Z}$ hierarchy.

\begin{figure}
\includegraphics[scale=0.5]{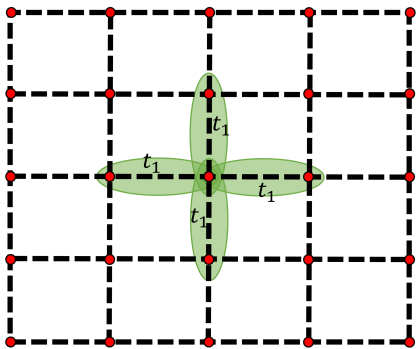}

\caption{\label{fig:Schematic}Representation of a particle in the central
site of a square lattice. Due to non-local terms of the Hamiltonian
(either nearest neighbor hopping $t_{1}$ or interaction $V_{1}$),
the particle becomes correlated with the neighboring sites. The importance
of these correlations is characterized by $\rho_{\mathbf{x},\mathbf{y}}^{C}$,
being $\mathbf{x},\mathbf{y}$ nearest neighbors. If the state is
strongly correlated, $\rho_{\mathbf{x},\mathbf{y}}^{C}$ has reached
a maximum value, and the entanglement with an extra site requires
that $\rho_{\mathbf{x},\mathbf{y}}^{C}$ must decrease as $1/\mathcal{Z}$.}
\end{figure}

Clearly one needs to justify the scaling $\rho_{\mathbf{x},\mathbf{y}}^{C}\sim\mathcal{Z}^{-1}$,
and prove that the scaling $\rho_{\mathbf{x},\mathbf{y}}^{C}\sim\mathcal{Z}^{-1}$
implies $V\sim\mathcal{Z}^{-1}$ for the interaction term (which also
proves the equivalence between the $\mathcal{Z}^{-1}$ hierarchy and
a hierarchy of correlations). To the question: Why is the scaling
of correlations inversely proportional to the coordination number
of the lattice?, one can argue that it must be related with the monogamy
property of entanglement\cite{Concurrence}: The monogamy of entanglement
states that a fully correlated state of particles cannot be entangled
with an extra particle, unless the entanglement between the initial
set is reduced. Let us apply this principle to a simple example: Consider
Fig.\ref{fig:Schematic}, where we have a simple square lattice populated
by spinless fermions. For simplicity, we assume that they can only
hop and interact with their nearest neighbors. Then, a simple Hamiltonian
describing the system would be:
\begin{equation}
H=-\sum_{\langle\mathbf{x},\mathbf{y}\rangle}t_{\mathbf{x},\mathbf{y}}\left(f_{\mathbf{y}}^{\dagger}f_{\mathbf{x}}+f_{\mathbf{x}}^{\dagger}f_{\mathbf{y}}\right)+\sum_{\langle\mathbf{x},\mathbf{y}\rangle}V_{\mathbf{x},\mathbf{y}}n_{\mathbf{x}}n_{\mathbf{y}}
\end{equation}
where $n_{\mathbf{x}}=f_{\mathbf{x}}^{\dagger}f_{\mathbf{x}}$ is
the number of particles at site $\mathbf{x}$, $t_{\mathbf{x},\mathbf{y}}$
is the hopping between sites and $V_{\mathbf{x},\mathbf{y}}$ corresponds
to the interaction between particles. In this representation, operators
are defined at different points of the lattice, and the hopping as
well as the interaction create correlations between sites, because
their value depends on the relative distance between pairs of points.
This is what we define as non-local terms of the Hamiltonian.

Let us calculate the time evolution of the time-ordered 2-point function
$G_{\mathbf{z},\mathbf{z}}^{T}\left(t,t^{\prime}\right)=-i\langle f_{\mathbf{z}}\left(t\right),f_{\mathbf{z}}^{\dagger}\left(t^{\prime}\right)\rangle_{T}$
using the Heisenberg's equation of motion, which corresponds to the
propagation of a particle excitation to and from site $\mathbf{z}$
(i.e., is proportional to $\rho_{\mathbf{z}}$):
\begin{eqnarray}
\partial_{t}G_{\mathbf{z},\mathbf{z}}^{T}\left(t,t^{\prime}\right) & = & -i\delta\left(t-t^{\prime}\right)+2i\sum_{\mathbf{x}\neq\mathbf{z}}t_{\mathbf{z},\mathbf{x}}G_{\mathbf{x},\mathbf{z}}^{T}\left(t,t^{\prime}\right)\nonumber \\
 &  & -2i\sum_{\mathbf{x}\neq\mathbf{z}}V_{\mathbf{z},\mathbf{x}}G_{\mathbf{x}\mathbf{x}\mathbf{z},\mathbf{z}}^{T}\left(t,t^{\prime}\right)
\end{eqnarray}
where we have defined $G_{\mathbf{x}\mathbf{x}\mathbf{z},\mathbf{z}}^{T}\left(t,t^{\prime}\right)=-i\langle n_{\mathbf{x}}\left(t\right)f_{\mathbf{z}}\left(t\right),f_{\mathbf{z}}^{\dagger}\left(t^{\prime}\right)\rangle_{T}$.
As the Green's function is calculated with respect to the density
matrix $\rho$, we can use the decomposition in terms of correlations
(Eq.\ref{eq:Decomposition}):
\begin{eqnarray}
\partial_{t}G_{\mathbf{z},\mathbf{z}}^{T}\left(t,t^{\prime}\right) & = & -i\delta\left(t-t^{\prime}\right)+2i\sum_{\mathbf{x}\neq\mathbf{z}}t_{\mathbf{z},\mathbf{x}}\mathcal{G}_{\mathbf{x},\mathbf{z}}^{T}\left(t,t^{\prime}\right)\\
 &  & -2i\sum_{\mathbf{x}\neq\mathbf{z}}V_{\mathbf{z},\mathbf{x}}\left[\langle n_{\mathbf{x}}\rangle G_{\mathbf{z},\mathbf{z}}^{T}\left(t,t^{\prime}\right)+\mathcal{G}_{\mathbf{x}\mathbf{x}\mathbf{z},\mathbf{z}}^{T}\left(t,t^{\prime}\right)\right]\nonumber 
\end{eqnarray}
where $\mathcal{G}$ denotes the statistical average with respect
to the correlated part of the density matrix $\rho_{\mathbf{x},\mathbf{z}}^{C}$.
If the Hamiltonian only contains local contributions, we would not
find terms proportional to $\mathcal{G}$, as correlations would not
propagate between different points of the lattice. However, the presence
of these terms induce correlations between modes, proportional to
the correlated part of the density matrix $\rho_{\mathbf{z},\mathbf{x}}^{C}$,
and the possibility of collective excitations. If we assume that the
resulting state between the particle excitation at site $\mathbf{z}$
and its nearest neighbors is a strongly correlated one, this implies
that $\rho_{\mathbf{z},\mathbf{x}}^{C}$ has reached a maximum value.
Therefore, due to the monogamy property of entanglement, a change
in the lattice including an extra nearest neighbor, would decrease
$\rho_{\mathbf{z},\mathbf{x}}^{C}$ as $\mathcal{Z}^{-1}$. This analysis
shows that the $\mathcal{Z}^{-1}$ scaling of correlations follows
directly from the monogamy of entanglement, if the system is strongly
correlated.

Finally, once we have physically motivated the scaling $\rho_{\mathbf{z},\mathbf{x}}^{C}\sim\mathcal{Z}^{-1}$,
we just need to relate the scaling of correlations with the physical
parameters of the Hamiltonian. This can be easily done by direct calculation
of the equation of motion for $\rho_{\mathbf{z},\mathbf{x}}^{C}$\cite{Queisser}.
One finds that in general, to lowest order in correlations, $\rho_{\mathbf{z},\mathbf{x}}^{C}$
turns out to be proportional to the non-local terms of the Hamiltonian
$t$ or $V$, times the product of the local density matrices $\rho_{\mathbf{x}}\rho_{\mathbf{z}}$.
This implies that the non-local terms of the Hamiltonian must scale
as $V,t\sim\mathcal{Z}^{-1}$. It is important to differenciate the
meaning of the scaling $\mathcal{Z}^{-1}$ from the actual value of
the interaction/hopping, which is fixed when the model is defined.
The meaning is that contributions to the statistical averages due
to entanglement between modes is reduced as $\mathcal{Z}^{-1}$, and
not that interactions become weaker as the coordination number increases.
To stress this feature, the $\mathcal{Z}^{-1}$ scaling is not directly
incorporated in the Hamiltonian.

In summary, we have shown that the monogamy property of entanglement
and the assumption of a strongly correlated state, naturally lead
to the $1/\mathcal{Z}$ scaling of correlations and of non-local terms
in the Hamiltonian. Furthermore, as spatial correlations can be extremely
different from the ones in $\mathbf{k}$ space, we will explore these
two different possibilities in the next sections and compare their
scaling properties for different models\footnote{We mention that while finalizing this manuscript we became aware of
a work with an analysis of the relation between the $1/\mathcal{Z}$
and the entanglement monogamy in terms of concurrence\cite{Entang-monogamy}.
We find their results consistent with the intuitive arguments given
in this work.}.

\section{Real space hierarchy}

As we discussed in the previous section, the hierarchy of correlations
can be established for different representations of the Hamiltonian.
In position space, correlations between different points are created
due to the non-local terms of the Hamiltonian. Furthermore, their
scaling is proportional to the inverse of the coordination number
$\mathcal{Z}$. Here we study the position space representation of
the hierarchy, for a system of spinless fermions in a lattice with
sub-lattice symmetry. The Hamiltonian can be written as:
\begin{eqnarray}
H & = & -\sum_{\mathbf{x},\mathbf{x}^{\prime}}\sum_{\sigma,\sigma^{\prime}}\left(J_{\mathbf{x},\mathbf{x}^{\prime}}^{\sigma,\sigma^{\prime}}f_{\mathbf{x},\sigma}^{\dagger}f_{\mathbf{x}^{\prime},\sigma^{\prime}}+\textrm{h.c.}\right)\label{eq:H1}\\
 &  & -\mu\sum_{\mathbf{x},\sigma}n_{\mathbf{x},\sigma}+\sum_{\mathbf{x},\mathbf{x}^{\prime}}\sum_{\sigma,\sigma^{\prime}}V_{\mathbf{x},\mathbf{x}^{\prime}}^{\sigma,\sigma^{\prime}}n_{\mathbf{x},\sigma}n_{\mathbf{x}^{\prime},\sigma^{\prime}}\nonumber 
\end{eqnarray}
where $f_{\mathbf{x},\sigma}^{\dagger}$ creates a fermion at position
$\mathbf{x}$ in sub-lattice $\sigma$ (the sub-index $\sigma$ can
characterize other internal degrees of freedom as well), and $n_{\mathbf{x},\sigma}=f_{\mathbf{x},\sigma}^{\dagger}f_{\mathbf{x},\sigma}$
is the number of particles operator. This Hamiltonian contains a hopping
term $J_{\mathbf{x},\mathbf{x}^{\prime}}^{\sigma,\sigma^{\prime}}$,
a chemical potential $\mu$ that fixes the number of particles in
the lattice, and a density-density interaction between the particles
$V_{\mathbf{x},\mathbf{x}^{\prime}}^{\sigma,\sigma^{\prime}}$.

The numerical calculations in this work are particularized for a dimerized
chain and a honeycomb lattice. For simplicity of our analysis we restrict
the models to nearest neighbors, although the analytical results are
valid for arbitrary number of neighbors interaction. We comment that
the use of short-range interactions for the honeycomb lattice corresponds
to a crude approximation of the real system due to the vanishing density
of states at zero filling. However, it is also a very interesting
approximation to understand the nature of phase transitions in 2D
models. The hopping and interaction terms are:
\begin{itemize}
\item \uline{Dimerized chain}: $J_{\mathbf{x},\mathbf{x}^{\prime}}^{\sigma,\sigma^{\prime}}=\delta_{\sigma^{\prime},\bar{\sigma}}\left(t_{1}\delta_{x,x^{\prime}}+t_{1}^{\prime}\delta_{x,x^{\prime}-a}\right)$
and $V_{\mathbf{x},\mathbf{x}^{\prime}}^{\sigma,\sigma^{\prime}}=V_{1}\delta_{\sigma^{\prime},\bar{\sigma}}\left(\delta_{x,x^{\prime}}+\delta_{x,x^{\prime}-a}\right)+V_{2}\delta_{\sigma^{\prime},\sigma}\delta_{x,x^{\prime}-a}$,
where for simplicity we have assumed that the interaction is symmetric
between the sub-lattices, $a$ is the length of the unit lattice vector
and $\bar{\sigma}=-\sigma$.
\item \uline{Honeycomb lattice}: $J_{\mathbf{x},\mathbf{x}^{\prime}}^{\sigma,\sigma^{\prime}}=t_{1}\delta_{\sigma^{\prime},\bar{\sigma}}\sum_{i}\delta_{\mathbf{x},\mathbf{x}^{\prime}+\mathbf{a}_{i}}$
and $V_{\mathbf{x},\mathbf{x}^{\prime}}^{\sigma,\sigma^{\prime}}=V_{1}\delta_{\sigma^{\prime},\bar{\sigma}}\sum_{i}\delta_{\mathbf{x},\mathbf{x}^{\prime}+\mathbf{a}_{i}}+V_{2}\delta_{\sigma^{\prime},\sigma}\sum_{i}\delta_{\mathbf{x},\mathbf{x}^{\prime}+\mathbf{b}_{i}}$,
where $\mathbf{a}_{i=1,2,3}$ ($\mathbf{b}_{i=1,\ldots,6}$) label
the lattice vectors connecting the different nearest (next nearest)
neighbors in the honeycomb lattice.
\end{itemize}
For the characterization of the physical properties we define a set
of double time Green's functions:
\begin{equation}
G_{\mathbf{y},\mathbf{y}^{\prime}}^{\alpha,\beta}\left(t,t^{\prime}\right)=-i\langle\langle f_{\mathbf{y},\alpha}\left(t\right);f_{\mathbf{y}^{\prime},\beta}^{\dagger}\left(t^{\prime}\right)\rangle\rangle
\end{equation}
where $\langle\langle\ldots;\ldots\rangle\rangle=\textrm{Tr}\left\{ \rho\ldots\right\} $
corresponds to the statistical average with respect to the density
matrix $\rho$, for the advanced/retarded/time ordered Green's function
(they all fulfill the same equation of motion, but with different
boundary conditions). We assume that the system is in thermal equilibrium,
and then the statistical average is calculated with respect to the
thermal density matrix $\rho=e^{-\beta H}$. If we make use of the
Heisenberg's equation of motion, and the fact that the system is in
thermodynamic equilibrium, we find that the 2-point Green's function
satisfies:
\begin{eqnarray}
\left(\omega+\mu\right)G_{\mathbf{y},\mathbf{y}^{\prime}}^{\alpha,\beta}\left(\omega\right) & = & \frac{\delta_{\mathbf{y},\mathbf{y}^{\prime}}\delta_{\alpha,\beta}}{2\pi}-2\sum_{\sigma,\mathbf{x}}J_{\mathbf{y},\mathbf{x}}^{\alpha,\sigma}G_{\mathbf{x},\mathbf{y}^{\prime}}^{\sigma,\beta}\left(\omega\right)\nonumber \\
 &  & +2\sum_{\sigma,\mathbf{x}}V_{\mathbf{y},\mathbf{x}}^{\alpha,\sigma}G_{\mathbf{x}\mathbf{x}\mathbf{y},\mathbf{y}^{\prime}}^{\sigma\sigma\alpha,\beta}\left(\omega\right)\label{eq:EOM1}
\end{eqnarray}
where we have introduced the 4-point function $G_{\mathbf{x}\mathbf{x}\mathbf{y},\mathbf{y}^{\prime}}^{\sigma\sigma\alpha,\beta}\left(\omega\right)\equiv-i\langle\langle f_{\mathbf{x},\sigma}^{\dagger}f_{\mathbf{x},\sigma}f_{\mathbf{y},\alpha};f_{\mathbf{y}^{\prime},\beta}^{\dagger}\rangle\rangle$.

Eq.\ref{eq:EOM1} corresponds to the general expression for the equation
of motion of the 2-point Green's function, which couples to higher
$n$-point Green's functions and creates correlations between modes.
The different approaches to find a solution to Eq.\ref{eq:EOM1} consist
in different decoupling schemes to separate the higher $n$-point
Green's functions into lower ones. As anticipated, in this work we
make use of the hierarchy of correlations.

Before we continue, we comment on the hierarchy of correlations when
couplings beyond nearest neighbors are included. Although we have
derived the hierarchy for the simple case of nearest neighbors only,
we can generalize the result to arbitrary neighbors: We can always
decompose a general interaction into different contributions according
to the lattice symmetries, e.g., $V_{\delta}^{\alpha,\sigma}=V_{N}^{\alpha,\sigma}+V_{NN}^{\alpha,\sigma}+\ldots$,
where $N$ refers to nearest neighbors, $NN$ to next nearest neighbors,
etc. Each of these components generates a hierarchy of correlations
with scaling proportional to $1/\mathcal{Z}_{N}$ for nearest neighbors,
$1/\mathcal{Z}_{NN}$ for next nearest neighbors, etc. The idea is
to find a lower bound for the scaling, where $V_{i}/\mathcal{Z}_{i}<V_{j}/\mathcal{Z}_{j}$
for all $i\neq j$. In this work we assume that the lower bound to
the scaling is given by $\mathcal{Z}$, i.e., is a short range interaction
in the sense $V_{N}/\mathcal{Z}_{N}<V_{NN}/\mathcal{Z}_{NN}<\ldots$.
This directly applies to the case of a dimerized chain, as the hopping
to nearest neighbors can be separated in two different contributions.
For simplicity we consider $t_{1}\simeq t_{1}^{\prime}$, which assumes
a unique scaling for both bonds, but one could also define the lower
bound for the hierarchy in terms of the two hoppings.

Finally, we expect that this hierarchy should be able to predict phases
with enlarged unit cells due to interactions\cite{PhaseDiag-HC1,HC-Interactions1,ExactDiag-HC}.
Although this is not discussed in the present work, we think that
for higher orders of the hierarchy, Green's functions with different
$\mathbf{k}$ vectors will couple in the equation of motion, and could
predict the transition to phases with enlarged unit cells.

\subsection{Uncorrelated solution}

As we are dealing with a hierarchy of equations, we must systematically
solve for the different $1/\mathcal{Z}$ orders. Here we first find
the lowest order solutions for the Green's functions, i.e., when spatial
correlations are neglected. If we expand Eq.\ref{eq:EOM1} in correlations,
and neglect terms of order $1/\mathcal{Z}$ or higher, we find:
\begin{eqnarray}
\left(\omega+\mu\right)g_{\mathbf{y},\mathbf{y}}^{\alpha,\beta} & = & \frac{\delta_{\alpha,\beta}}{2\pi}+2\sum_{\sigma,\mathbf{x}\neq\mathbf{y}}\bar{n}_{\mathbf{x},\sigma}V_{\mathbf{y},\mathbf{x}}^{\alpha,\sigma}g_{\mathbf{y},\mathbf{y}}^{\alpha,\beta}\label{eq:EOM2}
\end{eqnarray}
where $g_{\mathbf{y},\mathbf{y}}^{\alpha,\beta}$ denotes the uncorrelated
Green's function (neglecting $1/\mathcal{Z}$ corrections), the $1/\mathcal{Z}$
factor in $V_{\mathbf{y},\mathbf{x}}^{\alpha,\sigma}$ is compensated
by the sum over $\mathbf{x}$, and $\bar{n}_{\mathbf{x},\sigma}=\langle f_{\mathbf{x},\sigma}^{\dagger}f_{\mathbf{x},\sigma}\rangle_{0}$
corresponds to the uncorrelated local density of particles at site
$\mathbf{x}$ (we use the sub-index $0$ to indicate that all correlations
are neglected in the statistical average). Note that the equation
of motion is similar to what is expected for a mean field theory in
the strong coupling limit. As one would expect, the off-diagonal Green
functions vanish, as they are proportional to $\rho_{\mathbf{y},\mathbf{y}^{\prime}}^{C}$.
The solution to Eq.\ref{eq:EOM2} is easily obtained:
\begin{equation}
g_{\mathbf{y},\mathbf{y}}^{\alpha,\beta}=\frac{\delta_{\alpha,\beta}}{2\pi\left(\omega+\mu-2\sum_{\sigma,\mathbf{x}}\bar{n}_{\mathbf{x},\sigma}V_{\mathbf{y},\mathbf{x}}^{\alpha,\sigma}\right)}\label{eq:GF1}
\end{equation}
where we have included the term $\bar{n}_{\mathbf{y},\sigma}V_{\mathbf{y},\mathbf{y}}^{\alpha,\sigma}$
in the sum, in order to simplify the expression (this term does not
change the result, as it is of order $1/\mathcal{Z}$). Importantly,
one must deal with the sum in the denominator, and in general, it
could depend on long range interactions or on complicated local density
distributions. One simple approximation is to assume that the number
of particles in each sub-lattice is spatially homogeneous, i.e., we
assume that $\langle f_{\mathbf{x},\sigma}^{\dagger}f_{\mathbf{x},\sigma}\rangle_{0}=\bar{n}_{\sigma}=\sum_{\mathbf{x}}\bar{n}_{\mathbf{x},\sigma}/N$
is $\mathbf{x}$ independent. In this case, the solution reduces to:
\begin{equation}
g_{\mathbf{y},\mathbf{y}}^{\alpha,\beta}=\frac{\delta_{\alpha,\beta}}{2\pi\left(\omega-\omega_{0,\alpha}\right)}
\end{equation}
where $V_{\mathbf{q}}^{\alpha,\sigma}=\sum_{\delta}V_{\delta}^{\alpha,\sigma}e^{-i\mathbf{q}\delta}$
is the Fourier transform of the interaction potential and $\omega_{0,\alpha}=2\sum_{\sigma}\bar{n}_{\sigma}V_{0}^{\alpha,\sigma}-\mu$.
This approximation is a reasonable one, as when correlations are neglected,
all particles couple to an average field.

At this level of approximation, the structure of the Green's function
exhibits a single pole at $\omega_{0,\alpha}$, which is proportional
to the electron-electron interaction and the density of particles
in each sub-lattice. This means that excitations correspond to adding
or removing a particle, which couples to a charged background. Nevertheless,
we do not know the density of particles in each sub-lattice $n_{\bar{\alpha}}$,
and for a complete solution we must determine this value self-consistently.
This is done in the next subsection. Finally, we Fourier transform
the previous equation of motion to $\mathbf{k}$ space, as we will
make use of the Green's function in Fourier space in the next sections.
The solution is simply:
\begin{equation}
g_{\mathbf{k}}^{\alpha,\beta}=\frac{N\delta_{\alpha,\beta}\delta\left(\mathbf{k}\right)}{2\pi\left(\omega-\omega_{0,\alpha}\right)}\label{eq:g-kspace}
\end{equation}
where once again, we have assumed that the average number of particles
is spatially homogeneous in each sub-lattice:$\langle f_{\mathbf{x},\sigma}^{\dagger}f_{\mathbf{x},\sigma}\rangle_{0}^{\mathbf{q}}=N\bar{n}_{\sigma}\delta\left(\mathbf{q}\right)$.
Note that we use the superscript $\mathbf{q}$ to indicate the Fourier
transformed of the statistical average.

\subsubsection*{Self-consistency equations}

In order to find the physical properties of the system, we must determine
the different statistical averages by means of self-consistently equations.
This can be done using the next expression, which relates the double-time
Green's function with the statistical average\cite{Zubarev}:
\begin{equation}
\langle f_{\mathbf{y},\beta}^{\dagger}f_{\mathbf{y},\alpha}\rangle_{0}=i\int\frac{g_{\mathbf{y},\mathbf{y}}^{\alpha,\beta}\left(\omega+i\epsilon\right)-g_{\mathbf{y},\mathbf{y}}^{\alpha,\beta}\left(\omega-i\epsilon\right)}{e^{\beta\omega}+1}d\omega
\end{equation}
Then, if we make use of the uncorrelated solutions found in Eq.\ref{eq:g-kspace},
the average number of particles is:
\begin{equation}
\langle f_{\mathbf{y},\beta}^{\dagger}f_{\mathbf{y},\alpha}\rangle_{0}^{\mathbf{k}}=N\frac{\delta_{\alpha,\beta}\delta\left(\mathbf{k}\right)}{e^{\beta\omega_{0,\alpha}}+1}\label{eq:Average0}
\end{equation}
Importantly, this expression depends on the chemical potential $\mu$
and we need an extra equation for the solution. As $\mu$ is related
with the number of particles in the system, the extra equation can
be obtained from fixing the total number of particles:
\begin{equation}
N\sum_{\alpha}\bar{n}_{\alpha}=\sum_{\mathbf{x},\alpha}\langle f_{\mathbf{x},\alpha}^{\dagger}f_{\mathbf{x},\alpha}\rangle
\end{equation}
where $\bar{n}_{\alpha}$ is the average number of particles in each
sub-lattice:
\begin{eqnarray}
\bar{n}_{\alpha} & = & \frac{1}{e^{\beta\omega_{0,\alpha}}+1}
\end{eqnarray}
Now consider the case of a system with $\mathbb{Z}_{2}$ sub-lattice
symmetry (as the honeycomb lattice or the dimerized chain) at half
filling. We obtain a system of three coupled equations:
\begin{eqnarray}
\bar{n}_{\pm} & = & \frac{1}{e^{\beta\omega_{0,\pm}}+1},\ \bar{n}_{+}+\bar{n}_{-}=1
\end{eqnarray}
with three unknown quantities ($\bar{n}_{\pm}$ and $\mu$). We find
that the system of equations has a temperature independent solution
for $\mu=V_{0}^{+,+}+V_{0}^{+,-}$, and that the number of particles
is:
\begin{eqnarray}
\bar{n}_{\pm} & = & \frac{1}{e^{\pm\beta x\delta V_{0}}+1}
\end{eqnarray}
where $\delta V_{0}=V_{0}^{+,+}-V_{0}^{-,+}$, and $x=\bar{n}_{+}-\bar{n}_{-}$
is the asymmetry factor. As the self-consistency equation for the
asymmetry factor can be determined as well, we find:
\begin{equation}
x=-\tanh\left(\frac{\beta x\delta V_{0}}{2}\right)\label{eq:SCEq0}
\end{equation}
Clearly the solution $x=0$ is always valid, corresponding to a homogeneous
charge distribution on the lattice. In the homogeneous charge phase
(HCP) the particles are equally distributed between the two sub-lattices
and $\bar{n}_{\alpha}=1/2$. In addition, when nearest neighbor interaction
dominates over next nearest neighbor ($\delta V_{0}<0$), $x\neq0$
is also a solution with critical temperature given by $T_{c}=-\delta V_{0}$
(in units of the Boltzmann's constant $k_{B}$). In this case a charge
density wave (CDW) phase forms for temperatures lower than the critical
temperature. Note that when the system is at $T=0$ this means that
the CDW phase exists for any $V_{1}>0$. Thus, when correlations are
neglected, the statistical averages reveal an insulating Mott phase
with inhomogeneous charge distribution between the two sub-lattices.
Interestingly, the solution neglecting correlations coincides with
the strong coupling limit of a mean field theory.

\subsection{Correlated solution}

We now consider the full equation of motion for the Green's function
(Eq.\ref{eq:EOM1}), and include correlations up to $1/\mathcal{Z}$
order using the expansion in correlations:
\begin{eqnarray}
G_{\mathbf{x}\mathbf{x}\mathbf{y},\mathbf{y}^{\prime}}^{\sigma\sigma\alpha,\beta} & = & \bar{n}_{\mathbf{x},\sigma}G_{\mathbf{y},\mathbf{y}^{\prime}}^{\alpha,\beta}+\mathcal{G}_{\mathbf{x}\mathbf{x}\mathbf{y},\mathbf{y}^{\prime}}^{\sigma\sigma\alpha,\beta}
\end{eqnarray}
We find the next equations of motion:\begin{widetext}
\begin{eqnarray}
\left(\omega+\mu-2\sum_{\sigma,\mathbf{x}\neq\mathbf{y},\mathbf{y}^{\prime}}\bar{n}_{\mathbf{x},\sigma}V_{\mathbf{y},\mathbf{x}}^{\alpha,\sigma}\right)G_{\mathbf{y},\mathbf{y}}^{\alpha,\beta} & = & \frac{\delta_{\alpha,\beta}}{2\pi}-2\sum_{\sigma,\mathbf{x}\neq\mathbf{y}}J_{\mathbf{y},\mathbf{x}}^{\alpha,\sigma}\mathcal{G}_{\mathbf{x},\mathbf{y}}^{\sigma,\beta}-2\sum_{\sigma}J_{\mathbf{y},\mathbf{y}^{\prime}}^{\alpha,\sigma}g_{\mathbf{y},\mathbf{y}}^{\sigma,\beta}\nonumber \\
 &  & +2\sum_{\sigma,\mathbf{x}\neq\mathbf{y}}V_{\mathbf{y},\mathbf{x}}^{\alpha,\sigma}\mathcal{G}_{\mathbf{x}\mathbf{x}\mathbf{y},\mathbf{y}}^{\sigma\sigma\alpha,\beta}+2\sum_{\sigma}V_{\mathbf{y},\mathbf{y}}^{\alpha,\sigma}g_{\mathbf{y}\mathbf{y}\mathbf{y},\mathbf{y}}^{\sigma\sigma\alpha,\beta}\label{eq:EOM1-1}\\
\left(\omega+\mu-2\sum_{\sigma,\mathbf{x}\neq\mathbf{y},\mathbf{y}^{\prime}}\bar{n}_{\mathbf{x},\sigma}V_{\mathbf{y},\mathbf{x}}^{\alpha,\sigma}\right)\mathcal{G}_{\mathbf{y},\mathbf{y}^{\prime}}^{\alpha,\beta} & = & -2\sum_{\sigma,\mathbf{x}\neq\mathbf{y}^{\prime}}J_{\mathbf{y},\mathbf{x}}^{\alpha,\sigma}\mathcal{G}_{\mathbf{x},\mathbf{y}^{\prime}}^{\sigma,\beta}-2\sum_{\sigma}J_{\mathbf{y},\mathbf{y}^{\prime}}^{\alpha,\sigma}g_{\mathbf{y}^{\prime},\mathbf{y}^{\prime}}^{\sigma,\beta}\label{eq:EOM1-2}
\end{eqnarray}
\end{widetext}where we have separated diagonal ($\mathbf{y}=\mathbf{y}^{\prime}$)
and off-diagonal ($\mathbf{y}\neq\mathbf{y}^{\prime}$) Green's functions
due to their different scaling properties. We remind the reader that
the diagonal Green's function $G_{\mathbf{y},\mathbf{y}}^{\alpha,\beta}$
contains terms of order $1$ and $1/\mathcal{Z}$, $g_{\mathbf{y}\mathbf{y}\mathbf{y},\mathbf{y}}^{\sigma\sigma\alpha,\beta}\sim\mathcal{O}\left(1\right)$
corresponds to the uncorrelated part of the 4-point Green's function,
and $\mathcal{G}\sim\mathcal{O}\left(1/\mathcal{Z}\right)$ corresponds
to the correlated part of the Green's function.

Eq.\ref{eq:EOM1-1} contains contributions from the 4-point Green's
functions. For their calculation, we obtain the equation of motion
and expand the higher $n$-point Green's functions in correlations.
Neglecting all correlations we find that the uncorrelated 4-point
function is: 
\begin{eqnarray}
g_{\mathbf{y}\mathbf{y}\mathbf{y},\mathbf{y}}^{\sigma\sigma\alpha,\beta} & = & \langle f_{\mathbf{y},\sigma}^{\dagger}f_{\mathbf{y},\sigma}\rangle_{0}g_{\mathbf{y},\mathbf{y}}^{\alpha,\beta}-\langle f_{\mathbf{y},\sigma}^{\dagger}f_{\mathbf{y},\alpha}\rangle_{0}g_{\mathbf{y},\mathbf{y}}^{\sigma,\beta}\label{eq:Substitution1}
\end{eqnarray}
which turns out to be equivalent to use Wick's theorem. Note that
the pole structure is unchanged, as one would not expect new quasi-particles
for the description of the system in absence of correlations. In addition,
we have found that the expression for the 4-point function can be
written in terms of 2-point functions, and can be used to simplify
Eq.\ref{eq:EOM1-1}:
\begin{eqnarray}
\left(\omega-\omega_{0,\alpha}\right)G_{\mathbf{y},\mathbf{y}}^{\alpha,\beta} & = & \frac{\delta_{\alpha,\beta}}{2\pi}+2\sum_{\sigma,\mathbf{x}\neq\mathbf{y}}V_{\mathbf{y},\mathbf{x}}^{\alpha,\sigma}\mathcal{G}_{\mathbf{x}\mathbf{x}\mathbf{y},\mathbf{y}}^{\sigma\sigma\alpha,\beta}\label{eq:Full1}\\
 &  & -2\sum_{\sigma}\left(J_{\mathbf{y},\mathbf{y}}^{\alpha,\sigma}g_{\mathbf{y},\mathbf{y}}^{\sigma,\beta}+\sum_{\mathbf{x}\neq\mathbf{y}}J_{\mathbf{y},\mathbf{x}}^{\alpha,\sigma}\mathcal{G}_{\mathbf{x},\mathbf{y}}^{\sigma,\beta}\right)\nonumber 
\end{eqnarray}
where we have absorbed the first term of Eq.\ref{eq:Substitution1}
into $G_{\mathbf{y},\mathbf{y}^{\prime}}^{\alpha,\beta}\sum_{\mathbf{x}\neq\mathbf{y},\mathbf{y}^{\prime}}\bar{n}_{\mathbf{x},\sigma}V_{\mathbf{y},\mathbf{x}}^{\alpha,\sigma}$,
and used $\langle f_{\mathbf{y},\sigma}^{\dagger}f_{\mathbf{y},\alpha}\rangle_{0}\propto\delta_{\sigma,\alpha}$.
The Green's functions $\mathcal{G}_{\mathbf{x},\mathbf{y}^{\prime}}^{\sigma,\beta}$
and $\mathcal{G}_{\mathbf{x}\mathbf{x}\mathbf{y},\mathbf{y}}^{\sigma\sigma\alpha,\beta}$,
which are proportional to the correlated part of the density matrix,
are calculated from their equation of motion. We find next pair of
equations, which importantly, are decoupled from Eq.\ref{eq:Full1}:
\begin{eqnarray}
\left(\omega-\omega_{0,\alpha}\right)\mathcal{G}_{\mathbf{k},\mathbf{k}^{\prime}}^{\alpha,\beta} & = & -2J_{\mathbf{k}}^{\alpha,\beta}g_{\mathbf{k}+\mathbf{k}^{\prime}}^{\beta,\beta}-2J_{\mathbf{k}}^{\alpha,\bar{\alpha}}\mathcal{G}_{\mathbf{k},\mathbf{k}^{\prime}}^{\bar{\alpha},\beta}
\end{eqnarray}
\begin{eqnarray}
\left(\omega-\omega_{0,\alpha}\right)\mathcal{G}_{\mathbf{z}\mathbf{z}\mathbf{y},\mathbf{y}}^{\sigma\sigma\alpha,\beta} & = & 2\bar{n}_{\sigma}\left(1-\bar{n}_{\sigma}\right)g_{\mathbf{y},\mathbf{y}}^{\alpha,\beta}V_{\mathbf{y},\mathbf{z}}^{\alpha,\sigma}\\
 &  & +2g_{\mathbf{y},\mathbf{y}}^{\alpha,\beta}\sum_{\sigma^{\prime},\mathbf{x}\neq\mathbf{y},\mathbf{z}}V_{\mathbf{y},\mathbf{x}}^{\alpha,\sigma^{\prime}}\langle n_{\mathbf{z},\sigma}n_{\mathbf{x},\sigma^{\prime}}\rangle_{C}\nonumber 
\end{eqnarray}
being $\langle n_{\mathbf{z},\sigma}n_{\mathbf{x},\sigma^{\prime}}\rangle_{C}=\langle n_{\mathbf{z},\sigma}n_{\mathbf{x},\sigma^{\prime}}\rangle-\langle n_{\mathbf{z},\sigma}\rangle\langle n_{\mathbf{x},\sigma^{\prime}}\rangle$
the correlated part of the density-density correlation function. To
obtain these equations we have neglected next nearest neighbors hopping
(details of the calculation in the Appendix). The solutions after
a Fourier transform are given by:
\begin{equation}
\mathcal{G}_{\mathbf{k},\mathbf{k}^{\prime}}^{\alpha,\beta}=2g_{\mathbf{k}+\mathbf{k}^{\prime}}^{\beta,\beta}\frac{2\delta_{\alpha,\beta}\left|J_{\mathbf{k}}^{\alpha,\bar{\alpha}}\right|^{2}-\delta_{\bar{\alpha},\beta}J_{\mathbf{k}}^{\alpha,\beta}\left(\omega-\omega_{0,\beta}\right)}{\left(\omega-\omega_{\mathbf{k},+}\right)\left(\omega-\omega_{\mathbf{k},-}\right)}
\end{equation}
\begin{eqnarray}
\mathcal{G}_{\mathbf{k},\mathbf{k}^{\prime}}^{\sigma\sigma\alpha,\beta} & = & \delta_{\alpha,\beta}\frac{N\delta\left(\mathbf{k}+\mathbf{k}^{\prime}\right)\bar{n}_{\sigma}\left(1-\bar{n}_{\sigma}\right)V_{\mathbf{k}^{\prime}}^{\alpha,\sigma}}{\pi\left(\omega-\omega_{0,\alpha}\right)^{2}}\nonumber \\
 &  & +\delta_{\alpha,\beta}\frac{\sum_{\sigma^{\prime}}V_{\mathbf{k}^{\prime}}^{\alpha,\sigma^{\prime}}\langle n_{\mathbf{k},\sigma}n_{\mathbf{k}^{\prime},\sigma^{\prime}}\rangle_{C}}{\pi\left(\omega-\omega_{0,\alpha}\right)^{2}}\label{eq:Density-Density}
\end{eqnarray}
where the new poles of the Green's function $\mathcal{G}_{\mathbf{k},\mathbf{k}^{\prime}}^{\alpha,\beta}\left(\omega\right)$
are:
\begin{equation}
\omega_{\mathbf{k},\pm}=\pm\tilde{\omega}_{\mathbf{k}}=\pm\sqrt{4\left|J_{\mathbf{k}}^{-,+}\right|^{2}+x^{2}\delta V_{0}^{2}}\label{eq:poles2}
\end{equation}
$V_{0}=V_{0}^{+,+}+V_{0}^{-,+}$, $\delta V_{0}=V_{0}^{+,+}-V_{0}^{-,+}$
and we have fixed $\mu=V_{0}$. Finally, we can characterize the diagonal
Green's function including $1/\mathcal{Z}$ corrections (Eq.\ref{eq:Full1}):
\begin{eqnarray}
G_{\mathbf{k}}^{\alpha,\beta} & = & g_{\mathbf{k}}^{\alpha,\beta}-\frac{2}{N}\sum_{\mathbf{q}}J_{\mathbf{q}}^{\alpha,\bar{\alpha}}\frac{\mathcal{G}_{\mathbf{q},\mathbf{k}-\mathbf{q}}^{\bar{\alpha},\beta}}{\left(\omega-\omega_{0,\alpha}\right)}\label{eq:Full2}\\
 &  & -\frac{2J_{1}g_{\mathbf{k}}^{\bar{\alpha},\beta}}{\left(\omega-\omega_{0,\alpha}\right)}+\frac{2}{N}\sum_{\sigma,\mathbf{q}}V_{\mathbf{q}}^{\alpha,\sigma}\frac{\mathcal{G}_{\mathbf{q},\mathbf{k}-\mathbf{q}}^{\sigma\sigma\alpha,\beta}}{\left(\omega-\omega_{0,\alpha}\right)}\nonumber 
\end{eqnarray}
Note that in general we should extract the value of the chemical potential
from the self-consistency equations, once we have fixed the total
number of particles. Instead we will prove that our choice is correct
by checking with the self-consistency equation later on. Nevertheless,
this is the expected value when next nearest neighbor hopping is neglected.
The reason is that next nearest neighbor hopping shifts the conduction
and valence band, breaking particle-hole symmetry. This implies that
for half filling the chemical potential shifts from the neutral point.
Then if $J_{\mathbf{k}}^{\alpha,\alpha}$ is neglected, particle-hole
symmetry is recovered and the chemical potential is located between
the bands.

As one can observe in Eq.\ref{eq:Full2}, the Green's function $G_{\mathbf{k}}^{\alpha,\beta}$
contains several new contributions: The first term corresponds to
the uncorrelated solution previously found; the second term corresponds
to the hybridization between sites due to the hopping and it is characterized
by dispersive quasi-particle excitations; the third term only contributes
for $\alpha\neq\beta$, and represents a hopping between two different
sub-lattices, influenced by the density of particles in each of them;
the last term corresponds to the effect of density-density correlations.

\subsubsection*{Self-consistency equations}

\begin{figure}
\includegraphics[scale=0.65]{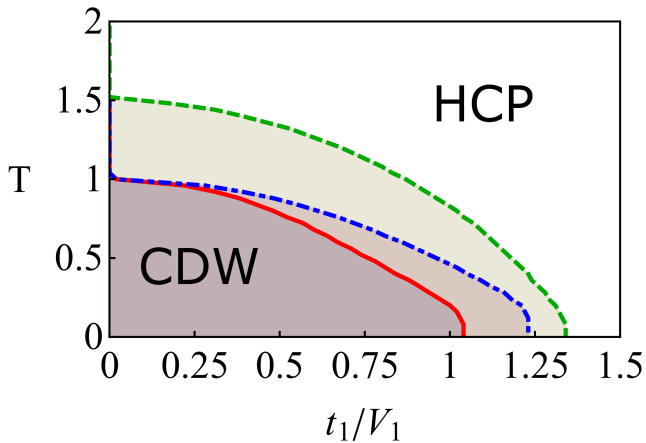}

\caption{\label{fig:Phase-diagram1}Phase diagram in absence of density-density
correlations for the real space hierarchy. We find a HCP and a CDW
phase as a function of $t_{1}/V_{1}$. We plot the case of a honeycomb
lattice (dashed green) and a dimerized chain in the topological ($\lambda=t_{1}^{\prime}/t_{1}=1.25$,
red) and trivial ($\lambda=0.75$, blue) phase, for $V_{2}=0$. The
addition of density-density correlations in the self-consistency equation
shifts the critical point to $t_{1}/V_{1}\rightarrow\infty$ (Fig.\ref{fig:Phase-diagram1-1}).}
\end{figure}

\begin{figure}
\includegraphics[scale=0.6]{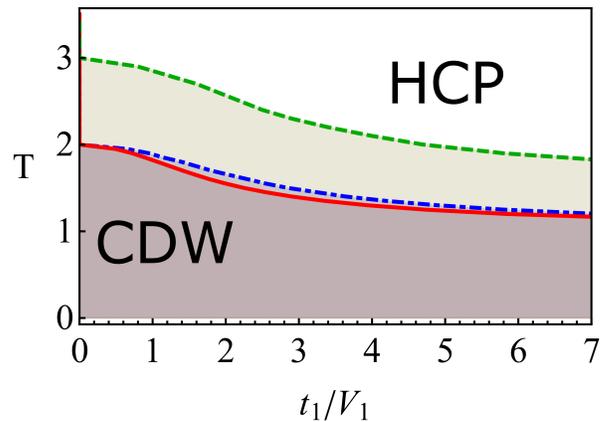}

\caption{\label{fig:Phase-diagram1-1}Full phase diagram for the real space
hierarchy including correlations ($V_{2}=0$). At $T=0$ the CDW phase
boundary extends to $t_{1}/V_{1}\rightarrow\infty$, meaning that
the HCP is only present in absence of interactions $V_{1}=0$, as
previously found in absence of correlations. Hybridization only affects
to the finite temperature region of the phase diagram, lowering the
critical temperature from $T_{C}=\mathcal{Z}V_{1}$.}
\end{figure}

We can use the previous expression for the 2-point function (Eq.\ref{eq:Full2})
to calculate the average number of particles in each sub-lattice:
\begin{eqnarray}
\langle n_{\mathbf{x},\alpha}\rangle^{\mathbf{k}} & = & i\int\frac{G_{\mathbf{k}}^{\alpha,\beta}\left(\omega+i\epsilon\right)-G_{\mathbf{k}}^{\alpha,\beta}\left(\omega-i\epsilon\right)}{e^{\beta\omega}+1}d\omega
\end{eqnarray}
In the Appendix we give details of the explicit calculation of the
statistical average. The general result is:
\begin{eqnarray}
\langle n_{\mathbf{x},\alpha}\rangle^{\mathbf{k}} & = & N\delta\left(\mathbf{k}\right)\left[\frac{1}{2}-\frac{1}{N}\sum_{\mathbf{q}}\frac{\omega_{0,\alpha}}{2\tilde{\omega}_{\mathbf{q}}}\tanh\left(\frac{\beta\tilde{\omega}_{\mathbf{q}}}{2}\right)\right]\label{eq:Number2}\\
 &  & -N\delta\left(\mathbf{k}\right)\bar{n}_{\alpha}\left(1-\bar{n}_{\alpha}\right)\textrm{sinh}\left(\beta\omega_{0,\alpha}\right)\nonumber 
\end{eqnarray}
where the last term corresponds to the contribution from density-density
correlations. It is clear from Eq.\ref{eq:Number2} that our choice
of the chemical potential ($\mu=V_{0}$) fulfills the self consistency
equation for the number of particles at half filling $\sum_{\mathbf{y},\alpha}\langle f_{\mathbf{y},\alpha}^{\dagger}f_{\mathbf{y},\alpha}\rangle/N=1$.
In addition, the self-consistency equation for the asymmetry factor
becomes:
\begin{equation}
x=-\frac{1}{N}\sum_{\mathbf{q}}\frac{x\delta V_{0}}{\tilde{\omega}_{\mathbf{q}}}\tanh\left(\frac{\beta\tilde{\omega}_{\mathbf{q}}}{2}\right)-\frac{1-x^{2}}{2}\textrm{sinh}\left(\beta x\delta V_{0}\right)\label{eq:SCEq1}
\end{equation}
Eq.\ref{eq:SCEq1} shows two contributions: On one hand, the first
term gives rise to a phase boundary between a homogeneous charge phase
($x=0$) and a CDW phase with broken sub-lattice symmetry ($x\neq0$),
characteristic of the RPA approximation (Fig.\ref{fig:Phase-diagram1});
the second term encodes the effect of density-density correlations,
which favor the CDW phase and dominate at low temperature (Fig.\ref{fig:Phase-diagram1-1}).
Importantly, the symmetry broken phase only exists for the case $\delta V_{0}=V_{0}^{+,+}-V_{0}^{+,-}<0$,
i.e., when nearest neighbors interaction $V_{1}$ dominates over next
nearest neighbors $V_{2}$. Note that the sum of both contributions
give rise to the critical temperature found in the last section $T_{c}=\mathcal{Z}V_{1}$
(for $t_{1}=0$), but the $T=0$ phase transition is lifted by the
second term to $t_{1}/V_{1}\rightarrow\infty$.

Therefore, density-density correlations stabilize the CDW phase, and
hybridization only affects the finite temperature phase boundary.
However, in contrast with the uncorrelated case, the system now contains
dispersive particle excitations (Eq.\ref{eq:poles2}) and the homogeneous
charge phase is not equivalent to the Mott insulator, previously found
in absence of correlations. For the case of the honeycomb lattice,
the homogeneous charge phase describes a semi-metal with two Dirac
cones.

We must comment that the finite temperature phase transitions in 1D
are artifacts coming from the mean-field-like approach to lowest order,
as in general the phase transitions should strictly exist at $T=0$
only. Furthermore, the presence of a CDW phase for infinitesimal interaction
goes against the results obtained from bosonization in 1D and well
established results in the honeycomb lattice. For these reason we
study in the next section an alternative form of the hierarchy for
$\mathbf{k}$ space correlations.

Finally, note that in the system with periodic boundary conditions,
the case $\lambda>1$ and $\lambda<1$ must be connected by a translation.
This can be easily seen from the self-consistency equation for $x$,
but in Fig.\ref{fig:Phase-diagram1} one must perform a re-scaling
$t_{1}\leftrightarrow t_{1}^{\prime}$ to confirm it.

\section{$\mathbf{k}$ Space hierarchy}

In this section we study the hierarchy for $\mathbf{k}$ space correlations.
Now the ground state is made of $\mathbf{k}$ space modes $\left\{ f_{\mathbf{k},\sigma}^{\dagger},f_{\mathbf{k},\sigma}\right\} $
initially decoupled, and the non-local terms introduce correlations
between them. If we Fourier transform Eq.\ref{eq:H1}, we obtain the
$\mathbf{k}$ space representation of the Hamiltonian:
\begin{eqnarray}
H & = & -\sum_{\begin{array}{c}
\mathbf{k},\sigma,\sigma^{\prime}\end{array}}\left(J_{\mathbf{k}}^{\sigma,\sigma^{\prime}}f_{\mathbf{k},\sigma}^{\dagger}f_{\mathbf{k},\sigma^{\prime}}+\textrm{h.c.}\right)-\mu\sum_{\mathbf{k},\sigma}n_{\mathbf{k},\sigma}\nonumber \\
 &  & +\frac{1}{N}\sum_{\mathbf{q}}\sum_{\begin{array}{c}
\mathbf{k},\mathbf{k}^{\prime}\\
\sigma,\sigma^{\prime}
\end{array}}V_{\mathbf{q}}^{\sigma,\sigma^{\prime}}f_{\mathbf{k},\sigma}^{\dagger}f_{\mathbf{k}^{\prime},\sigma^{\prime}}^{\dagger}f_{\mathbf{k}^{\prime}-\mathbf{q},\sigma^{\prime}}f_{\mathbf{k}+\mathbf{q},\sigma}\label{eq:H2}
\end{eqnarray}
where we have used:
\begin{eqnarray}
f_{\mathbf{x},\sigma}^{\dagger} & = & \frac{1}{\sqrt{N}}\sum_{\mathbf{k}}f_{\mathbf{k},\sigma}^{\dagger}e^{i\mathbf{k}\cdot\mathbf{x}}\\
f_{\mathbf{x},\sigma} & = & \frac{1}{\sqrt{N}}\sum_{\mathbf{k}}f_{\mathbf{k},\sigma}e^{-i\mathbf{k}\cdot\mathbf{x}}
\end{eqnarray}
As for the real space hierarchy, we need to characterize the scaling
properties of correlations in $\mathbf{k}$ space, which in general
will be different from the scaling of spatial correlations. We apply
a similar analysis to the one used in the first section of this manuscript,
studying the propagation of correlations for an excitation created
with some initial momentum $\mathbf{k}$. However, let us first discuss
some properties of Eq.\ref{eq:H2}: One important difference with
the real space representation (Eq.\ref{eq:H1}) is that the hopping
is now a local term (the momentum $\mathbf{k}$ of an excitation does
not change if we neglect particle interactions). Furthermore, the
behavior of $V_{\mathbf{q}}$ as a function of $\mathbf{q}$ is completely
different from the one of $V_{\mathbf{x}}$ as a function of $\mathbf{x}$,
which is a crucial change with respect to the real space hierarchy.
The reason is that short range potentials become long range in $\mathbf{k}$
space. Then, we expect correlations to spread differently, and for
some potentials, between a larger number of modes than in real space.

To define the scaling in a more precise way and study the system properties,
we introduce the next 2-point function:
\begin{equation}
G_{\mathbf{k}}^{\alpha,\beta}\left(t,t^{\prime}\right)=-i\langle\langle f_{\mathbf{k},\alpha}\left(t\right);f_{\mathbf{k},\beta}^{\dagger}\left(t^{\prime}\right)\rangle\rangle
\end{equation}
which characterizes the propagation of a particle with momentum $\mathbf{k}$
from sub-lattice $\beta$ to $\alpha$. As in the previous section,
we calculate the equation of motion:
\begin{eqnarray}
\left(\omega+\mu\right)G_{\mathbf{k}}^{\alpha,\beta} & = & \frac{\delta_{\alpha,\beta}}{2\pi}-2\sum_{\sigma}J_{\mathbf{k}}^{\alpha,\sigma}G_{\mathbf{k}}^{\sigma,\beta}\label{eq:EOM-k-1}\\
 &  & +\frac{2}{N}\sum_{\mathbf{k}_{1},\mathbf{q},\sigma}V_{\mathbf{q}}^{\alpha,\sigma}G_{\mathbf{k}_{1},\mathbf{k}_{1}-\mathbf{q},\mathbf{k}+\mathbf{q};\mathbf{k}}^{\sigma,\sigma,\alpha;\beta}\nonumber 
\end{eqnarray}
where we have defined the 4-point Green's function $G_{\mathbf{k}_{1},\mathbf{k}_{1}-\mathbf{q},\mathbf{k}+\mathbf{q};\mathbf{k}}^{\sigma,\sigma,\alpha;\beta}=-i\langle\langle f_{\mathbf{k}_{1},\sigma}^{\dagger}f_{\mathbf{k}_{1}-\mathbf{q},\sigma}f_{\mathbf{k}+\mathbf{q},\alpha};f_{\mathbf{k},\beta}^{\dagger}\rangle\rangle$.
We see from Eq.\ref{eq:EOM-k-1} that due to the $\mathbf{q}$ dependence
of $V_{\mathbf{q}}$, the analysis is more involved than in the real
space hierarchy. Let us first expand to linear order in correlations
the non-local term in Eq.\ref{eq:EOM-k-1} (i.e., we include up to
two point correlations):\begin{widetext}
\begin{eqnarray}
\sum_{\mathbf{k}_{1},\mathbf{q}}V_{\mathbf{q}}^{\alpha,\sigma}G_{\mathbf{k}_{1},\mathbf{k}_{1}-\mathbf{q},\mathbf{k}+\mathbf{q};\mathbf{k}}^{\sigma,\sigma,\alpha;\beta} & \simeq & \sum_{\mathbf{k}_{1}}V_{0}^{\alpha,\sigma}\bar{n}_{\mathbf{k}_{1},\sigma}G_{\mathbf{k}}^{\alpha,\beta}-\sum_{\mathbf{k}_{1}}V_{\mathbf{k}_{1}-\mathbf{k}}^{\alpha,\sigma}\langle f_{\mathbf{k}_{1},\sigma}^{\dagger}f_{\mathbf{k}_{1},\alpha}\rangle G_{\mathbf{k}}^{\sigma,\beta}\nonumber \\
 &  & +\sum_{\mathbf{k}_{1}}V_{0}^{\alpha,\sigma}\mathcal{G}_{\mathbf{k}_{1},\mathbf{k}_{1},\mathbf{k};\mathbf{k}}^{\sigma,\sigma,\alpha;\beta}-\sum_{\mathbf{k}_{1}}V_{\mathbf{k}_{1}-\mathbf{k}}^{\alpha,\sigma}\mathcal{G}_{\mathbf{k}_{1},\mathbf{k}_{1},\mathbf{k};\mathbf{k}}^{\sigma,\alpha,\sigma;\beta}\label{eq:K-Hierarchy}
\end{eqnarray}
\end{widetext}where we have neglected terms involving three or more
points (this is not important as they are expected to be of higher
order in the corresponding scaling parameter). Note that the two dominant
contributions at this order are $\mathbf{q}=0$ and $\mathbf{q}=\mathbf{k}_{1}-\mathbf{k}$,
which correspond to a direct and exchange channels, respectively.
The first line of Eq.\ref{eq:K-Hierarchy} contains the uncorrelated
contributions, while the second line the correlated ones. 

We discuss the term proportional to $V_{0}^{\alpha,\sigma}$ first:
We assume that the system is in a highly correlated state, which implies
that $\mathcal{G}_{\mathbf{k}_{1},\mathbf{k}_{1},\mathbf{k};\mathbf{k}}^{\sigma,\sigma,\alpha;\beta}$
has reached a maximum value as a function of $\rho_{\mathbf{k}_{1},\mathbf{k}}^{C}$.
If we include an additional particle to the lattice (i.e., an extra
state in the sum), we find that due to the entanglement monogamy $\mathcal{G}_{\mathbf{k}_{1},\mathbf{k}_{1},\mathbf{k};\mathbf{k}}^{\sigma,\sigma,\alpha;\beta}$
must decrease as $1/N_{p}$, being $N_{p}$ the number of particles
in the system. Therefore, the scaling of the non-local interaction
term $V_{\mathbf{q}=0}^{\alpha,\sigma}$ must be proportional to $1/N_{p}$.
Note that this only applies to correlations created by $V_{0}^{\alpha,\sigma}$,
and in order to find a lower bound for the scaling, we must analyze
$V_{\mathbf{q}\neq0}^{\alpha,\sigma}$ as well.

\begin{figure}
\includegraphics[scale=0.75]{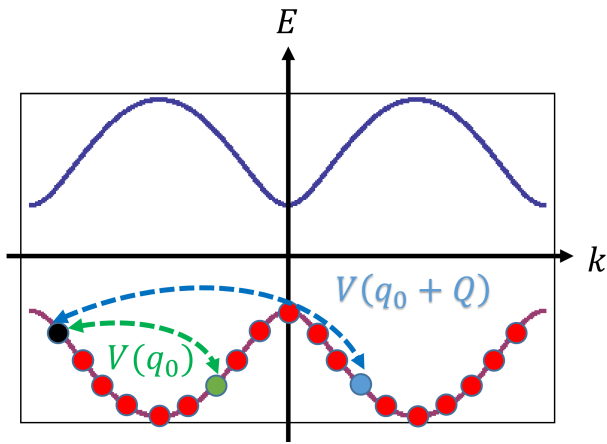}

\caption{\label{fig:K-space-correlations}Correlations in $\mathbf{k}$ space.
An initial state with momentum $\mathbf{k}$ interacts with all other
particles in the system via the exchange of $\mathbf{q}$. If we assume
for simplicity that some $\mathbf{q}_{0}$ dominates the interaction
process, we show for a section of the BZ that different states can
couple with the same $V_{\mathbf{q}_{0}}=V_{\mathbf{q}_{0}+n\mathbf{Q}}$,
as $V_{\mathbf{q}}$ is a periodic function of $\mathbf{q}$. Then
for a large system we find that $n\propto N_{p}$, and the scaling
of the interaction term $V_{\mathbf{q}_{0}}$ must be proportional
to $N_{p}^{-1}$.}
\end{figure}

As the case with $\mathbf{q}\neq0$ is more complicated, let us assume
that the interaction potential $V_{\mathbf{q}}$ dominates for some
$\mathbf{q}_{0}=\mathbf{k}_{1}-\mathbf{k}\neq0$. Due to the nearest
neighbors approximation used in the real space hierarchy, the interaction
potential $V_{\mathbf{q}}$ is a periodic function of $\mathbf{q}$,
i.e., $V_{\mathbf{q}_{0}}=V_{\mathbf{q}_{0}+\mathbf{Q}}$ for some
$\mathbf{Q}$ vector in the Brillouin zone (BZ). The systems considered
in this work are at half-filling and have sub-lattice symmetry, which
implies that every $\mathbf{k}_{1}$ point couples to all other points
in the BZ, as they are all occupied. Then for a large system we can
rewrite the sum as $\sum_{\mathbf{k}_{1}}V_{\mathbf{k}_{1}-\mathbf{k}}^{\alpha,\sigma}\mathcal{G}_{\mathbf{k}_{1},\mathbf{k}_{1},\mathbf{k};\mathbf{k}}^{\sigma,\alpha,\sigma;\beta}=V_{\mathbf{q}_{0}}^{\alpha,\sigma}\sum_{n}\mathcal{G}_{\mathbf{k}+n\mathbf{Q},\mathbf{k}+n\mathbf{Q},\mathbf{k};\mathbf{k}}^{\sigma,\alpha,\sigma;\beta}$
($n\in\mathbb{Z}$), where $n$ runs over a large number of values,
that for simplicity, we approximate as being proportional to $N_{p}$
(it will be smaller, but still a large number compared with $\mathcal{Z}$
in the real space hierarchy). As previously, if we add a new state
into the sum, the correlations between all previous particles must
decrease as $n^{-1}\propto N_{p}^{-1}$. As we assumed that $\mathbf{q}_{0}$
was the dominant contribution, all other $\mathbf{q}$ values should
provide a better scaling, and we can conclude that the lower bound
is given by $V_{\mathbf{q}}\propto N_{p}^{-1}$ for all $\mathbf{q}$.
In Fig.\ref{fig:K-space-correlations} we illustrate the scaling properties
of the interaction part of the Hamiltonian following the previous
discussion. Note that the number of points connected with the same
$V_{\mathbf{q}_{0}}$ correspond to the analog of the coordination
number in the real space hierarchy, and strongly depend on the shape
of the Brillouin zone and on the specific interaction potential. However,
many typical interactions considered in many-body systems would give
rise to a periodic $V_{\mathbf{q}}$, which provides a general improvement
over the $1/\mathcal{Z}$ scaling. More complicated interactions such
as dipolar must be carefully analyzed to find if the scaling of spatial
or $\mathbf{k}$ space correlations is more efficient\footnote{Importantly, note that for a system away from half-filling or without
sub-lattice symmetry the analysis of the scaling properties of correlations
is more complicated, as some $\mathbf{k}_{1}$ states will not be
occupied. This implies that a more asymmetric distribution of correlations
for different $\mathbf{k}$ states will be favored.}.

It is worth mentioning that the scaling of correlations in $\mathbf{k}$
space is, in many cases, more efficient than the real space one, specially
when we are dealing with low coordination numbers, as for the case
of low dimensional systems. Nevertheless, as we will determine the
self-consistency equations in this approach, we can directly compare
the results of both hierarchies.

We stress that the general arguments given in this manuscript to estimate
the scaling of correlations, rely on the assumption of having a fully
correlated state. In more detailed calculation the precise scaling
could be estimated in terms of direct calculations of entanglement
between modes. However, the arguments used in this manuscript allow
us to compare different hierarchies in a very systematic way, which
is the purpose of this work.

\subsection{Uncorrelated solution}

We first obtain the uncorrelated solutions to Eq.\ref{eq:EOM-k-1}.
If we make use of the scaling of the interaction term $V_{\mathbf{q}}\propto N_{p}^{-1}$,
we find:
\begin{eqnarray}
\left(\omega-\omega_{0,\alpha}\right)g_{\mathbf{k}}^{\alpha,\beta} & = & \frac{\delta_{\alpha,\beta}}{2\pi}-2\sum_{\sigma}J_{\mathbf{k}}^{\alpha,\sigma}g_{\mathbf{k}}^{\sigma,\beta}
\end{eqnarray}
where $\bar{n}_{\sigma}=\frac{1}{N}\sum_{\mathbf{k}_{1}}\bar{n}_{\mathbf{k}_{1},\sigma}$
is the average number of particles in each sub-lattice. As the system
of equations is diagonal in $\mathbf{k}$, the solution to the Green's
function is easily obtained:
\begin{eqnarray}
g_{\mathbf{k}}^{\alpha,\alpha} & = & \frac{\omega+\mu-2\sum_{\sigma}\bar{n}_{\sigma}V_{0}^{\bar{\alpha},\sigma}}{2\pi\left(\omega-\omega_{\mathbf{k},+}\right)\left(\omega-\omega_{\mathbf{k},-}\right)}\\
g_{\mathbf{k}}^{\alpha,\bar{\alpha}} & = & \frac{-J_{\mathbf{k}}^{\alpha,\bar{\alpha}}}{\pi\left(\omega-\omega_{\mathbf{k},+}\right)\left(\omega-\omega_{\mathbf{k},-}\right)}
\end{eqnarray}
where we have defined:
\begin{eqnarray}
\omega_{\mathbf{k},\pm} & = & \left(\bar{n}_{+}+\bar{n}_{-}\right)V_{0}-\mu\pm\tilde{\omega}_{\mathbf{k}}\\
\tilde{\omega}_{\mathbf{k}} & = & \sqrt{x^{2}\delta V_{0}^{2}+4\left|J_{\mathbf{k}}^{-,+}\right|^{2}}
\end{eqnarray}
and assumed nearest neighbors hopping only. Note that the Green's
functions are the ones expected from the RPA approximation, and the
quasi-particles are dispersive due to hybridization. The calculation
of the average number of particles $\bar{n}_{\mathbf{k},\alpha}$and
average hybridization $\Delta_{\mathbf{k}}^{\bar{\alpha},\alpha}$
gives:
\begin{eqnarray}
\bar{n}_{\mathbf{k},\alpha} & = & \frac{1}{2}-\alpha\frac{x\delta V_{0}}{2\tilde{\omega}_{\mathbf{k}}}\tanh\left(\frac{\beta\tilde{\omega}_{\mathbf{k}}}{2}\right)\\
\langle f_{\mathbf{k},\bar{\alpha}}^{\dagger}f_{\mathbf{k},\alpha}\rangle_{0} & = & \Delta_{\mathbf{k}}^{\bar{\alpha},\alpha}=\frac{J_{\mathbf{k}}^{\alpha,\bar{\alpha}}}{\tilde{\omega}_{\mathbf{k}}}\tanh\left(\frac{\beta\tilde{\omega}_{\mathbf{k}}}{2}\right)
\end{eqnarray}
Providing the next self-consistency equation for the asymmetry factor
$x$: 
\begin{eqnarray}
x & = & -\frac{1}{N}\sum_{\mathbf{k}}\frac{x\delta V_{0}}{\tilde{\omega}_{\mathbf{k}}}\tanh\left(\frac{\beta\tilde{\omega}_{\mathbf{k}}}{2}\right)\label{eq:SCE-Uncorrk}
\end{eqnarray}
where we have assumed half filling, and the chemical potential has
been fixed to $\mu=V_{0}$, which allows to fulfill the self-consistency
equation for the total number of particles, at arbitrary temperature.
The phase diagram corresponds to the one shown in Fig.\ref{fig:Phase-diagram1},
where the transition between the CDW and the homogeneous charge phase
is controlled by the ratio $\delta V_{0}/t_{1}$. It is interesting
that the self-consistency equation (Eq.\ref{eq:SCE-Uncorrk}), corresponds
to the one found in the real space hierarchy in absence of density-density
correlations (Eq.\ref{eq:SCEq1}). Therefore in this case, the homogeneous
charge phase is present for finite values of $V_{1}$ at $T=0$, as
shown in Fig.\ref{fig:Phase-diagram1}. In addition, the critical
temperature $T_{c}$ is half the value that was found in the real
space hierarchy. This result is in better agreement with the exact
diagonalization and functional renormalization group studies\cite{ExactDiag-HC,Scherer2015}
(both on the honeycomb lattice), which confirms that the $\mathbf{k}$
space hierarchy is more effective for the case of short ranged potentials.

\subsection{Correlated solution}

Here we include the effect of $\mathbf{k}$ space correlations. We
start from the full equation of motion for the 2-point Green's function
(Eq.\ref{eq:EOM-k-1}), and include terms of order $N_{p}^{-1}$.
We find:\begin{widetext}
\begin{eqnarray}
\left(\omega+\mu\right)G_{\mathbf{k}}^{\alpha,\beta} & = & \frac{\delta_{\alpha,\beta}}{2\pi}-2\sum_{\sigma}J_{\mathbf{k}}^{\alpha,\sigma}G_{\mathbf{k}}^{\sigma,\beta}-\frac{2}{N}\sum_{\sigma,\mathbf{q}\neq0}V_{\mathbf{q}}^{\sigma,\alpha}\langle f_{\mathbf{k}-\mathbf{q},\sigma}^{\dagger}f_{\mathbf{k}-\mathbf{q},\alpha}\rangle_{0}g_{\mathbf{k}}^{\sigma,\beta}\nonumber \\
 &  & +\frac{2}{N}G_{\mathbf{k}}^{\alpha,\beta}\sum_{\sigma,\mathbf{k}_{1}}\bar{n}_{\mathbf{k}_{1},\sigma}V_{0}^{\alpha,\sigma}+\frac{2}{N}\sum_{\sigma,\mathbf{k}_{1}\neq\mathbf{k}}V_{0}^{\alpha,\sigma}\mathcal{G}_{\mathbf{k}_{1},\mathbf{k}_{1},\mathbf{k};\mathbf{k}}^{\sigma,\sigma,\alpha;\beta}
\end{eqnarray}
\end{widetext}where we have separated the $\mathbf{q}=0$ and $\mathbf{k}_{1}=\mathbf{k}+\mathbf{q}$
terms in the sum, as they are the only ones that contribute to first
order. We also neglected the contribution from $V_{0}^{\alpha,\sigma}g_{\mathbf{k},\mathbf{k},\mathbf{k};\mathbf{k}}^{\sigma,\sigma,\alpha;\beta}$,
which scales inversely proportional to the volume of the system. The
contribution from the correlated part of the 4-point function $\mathcal{G}_{\mathbf{k}_{1},\mathbf{k}_{1},\mathbf{k};\mathbf{k}}^{\sigma,\sigma,\alpha;\beta}$
can be obtained in a similar way to the previous section. We have
calculated its equation of motion and found that in general, to first
order, the correlated part of the density-density correlations $\langle n_{\mathbf{k},\alpha}n_{\mathbf{k}^{\prime},\beta}\rangle_{C}$
can be neglected. Therefore the consequence of first order correlations
in the equation of motion is the presence of response functions (convolutions)
between the local statistical averages and the interaction potentials:
$V_{\mathbf{k}}^{\sigma,\alpha}\ast\langle f_{\mathbf{k},\sigma}^{\dagger}f_{\mathbf{k},\alpha}\rangle_{0}=\sum_{\mathbf{q}}V_{\mathbf{q}}^{\sigma,\alpha}\langle f_{\mathbf{k}-\mathbf{q},\sigma}^{\dagger}f_{\mathbf{k}-\mathbf{q},\alpha}\rangle_{0}/N$.

We rewrite the equation of motion in the next compact form:
\begin{eqnarray}
\left(\omega-\omega_{0,\alpha}\right)G_{\mathbf{k}}^{\alpha,\beta} & = & \frac{\delta_{\alpha,\beta}}{2\pi}-2\sum_{\sigma}J_{\mathbf{k}}^{\alpha,\sigma}G_{\mathbf{k}}^{\sigma,\beta}-2\Lambda_{\mathbf{k}}^{\alpha,\beta}
\end{eqnarray}
where $\Lambda_{\mathbf{k}}^{\alpha,\beta}=\sum_{\sigma}g_{\mathbf{k}}^{\sigma,\beta}V_{\mathbf{k}}^{\sigma,\alpha}\ast\langle f_{\mathbf{k},\sigma}^{\dagger}f_{\mathbf{k},\alpha}\rangle_{0}$
corresponds to the response functions. The general solution to the
system of equations is:
\begin{equation}
G_{\mathbf{k}}^{\alpha,\beta}=g_{\mathbf{k}}^{\alpha,\beta}+\frac{4J_{\mathbf{k}}^{\alpha,\bar{\alpha}}\Lambda_{\mathbf{k}}^{\bar{\alpha},\beta}-2\Lambda_{\mathbf{k}}^{\alpha,\beta}\left(\omega-\omega_{0,\bar{\alpha}}\right)}{\left(\omega-\omega_{\mathbf{k},+}\right)\left(\omega-\omega_{\mathbf{k},-}\right)}\label{eq:TotalGF}
\end{equation}
The Green's functions including correlations show the next interesting
features:
\begin{itemize}
\item As $\Lambda_{\mathbf{k}}^{\alpha,\beta}\propto g_{\mathbf{k}}^{\sigma,\beta}$,
the pole structure of the new contributions has order two poles. Naively
one could think that this is the result from simple perturbation theory,
however the presence of response functions proves otherwise. Furthermore,
the presence of poles of order two is related to the fact that for
the description of 2-point correlations we need pairs of excitations.
\item The effect of correlations is encoded in the response functions $\Lambda_{\mathbf{k}}^{\alpha,\beta}$
and they represent the effect of interactions in the density distribution.
As they are present in the numerator, they modify the polynomial structure
of the Green's functions allowing for the appearance of zeros. This
will be related with the topological properties in the next section.
\end{itemize}

\subsubsection*{Self-consistency equations}

\begin{figure}
\includegraphics[scale=0.6]{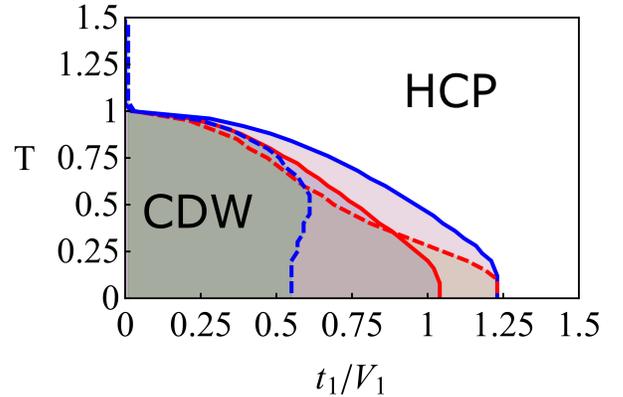}

\caption{\label{fig:Full-Phase-diagram}Phase diagram for the dimerized chain
with nearest neighbors hopping and interaction ($V_{2}=0$). We have
chosen $\lambda=1.25$ and $0.75$ ( red and blue color, respectively).
The continuous lines correspond to the uncorrelated case, while the
dashed lines correspond to the one including 2-point correlations.
It is interesting to see how correlations play an opposite role in
the stabilization of the different phases. While the HCP is stabilized
for the case $\lambda<1$ (blue), the CDW phase is stabilized for
$\lambda>1$ (red). The $T=0$ critical points shift to different
values due to correlations, however the critical temperature remains
unchanged. It is interesting to see that the correlated phase diagram
for the case $\lambda<1$ case shows a bump near $T\simeq0.4$, where
the CDW is stable for larger ratio $t_{1}/V_{1}$ than at the critical
point.}
\end{figure}

In this section we first write explicit expressions for the response
functions, and then we characterize the self-consistency equations
including correlations. We first assume that the chemical potential
is unchanged when correlations are included (this will be confirmed
by direct calculation of the self-consistency equation for the number
of particles), then we have $\mu=V_{0}^{+,+}+V_{0}^{-,+}$ for $J_{\mathbf{k}}^{\alpha,\alpha}=0$.
The expressions for the convolutions show that in this case, we have
the next relation between them:
\begin{eqnarray}
\chi_{\mathbf{k}}^{+,+} & = & -x\frac{\delta V_{0}}{2N}\sum_{\mathbf{q}}\frac{V_{\mathbf{q}}^{+,+}}{\tilde{\omega}_{\mathbf{k}-\mathbf{q}}}\tanh\left(\frac{\beta\tilde{\omega}_{\mathbf{k}-\mathbf{q}}}{2}\right)\nonumber \\
 & = & -\chi_{\mathbf{k}}^{-,-}\\
\chi_{\mathbf{k}}^{-,+} & = & \frac{1}{N}\sum_{\mathbf{q}}V_{\mathbf{q}}^{-,+}\frac{J_{\mathbf{k}-\mathbf{q}}^{+,-}}{\tilde{\omega}_{\mathbf{k}-\mathbf{q}}}\tanh\left(\frac{\beta\tilde{\omega}_{\mathbf{k}-\mathbf{q}}}{2}\right)\nonumber \\
 & = & \left(\chi_{\mathbf{k}}^{+,-}\right)^{\ast}
\end{eqnarray}
where we have used the expressions for the uncorrelated statistical
averages previously found, and the shorthand notation for the convolutions
$\chi_{\mathbf{k}}^{\sigma,\alpha}=V_{\mathbf{k}}^{\sigma,\alpha}\ast\langle f_{\mathbf{k},\sigma}^{\dagger}f_{\mathbf{k},\alpha}\rangle_{0}$.
Therefore, there are just two independent response functions. The
calculation of the statistical averages is now straight forward, as
the Green's functions highly simplify for our choice of the chemical
potential:
\begin{equation}
G_{\mathbf{k}}^{\alpha,\beta}=g_{\mathbf{k}}^{\alpha,\beta}+\frac{4J_{\mathbf{k}}^{\alpha,\bar{\alpha}}\Gamma_{\mathbf{k}}^{\bar{\alpha},\beta}-2\Gamma_{\mathbf{k}}^{\alpha,\beta}\left(\omega+\alpha x\delta V_{0}\right)}{\left(\omega-\omega_{\mathbf{k},+}\right)^{2}\left(\omega-\omega_{\mathbf{k},-}\right)^{2}}
\end{equation}
where $\Gamma_{\mathbf{k}}^{\alpha,\beta}$ corresponds to $\Lambda_{\mathbf{k}}^{\alpha,\beta}$
without the pole contribution. The expectation values including correlations
are given by:\begin{widetext}
\begin{eqnarray}
\langle n_{\mathbf{k},\pm}\rangle & = & \frac{1}{2}\pm\frac{2\chi_{\mathbf{k}}^{+,+}-x\delta V_{0}}{2\tilde{\omega}_{\mathbf{k}}}\tanh\left(\frac{\beta\tilde{\omega}_{\mathbf{k}}}{2}\right)\label{eq:Average-n}\\
 &  & \pm x\delta V_{0}\frac{J_{\mathbf{k}}^{-,+}\chi_{\mathbf{k}}^{-,+}+J_{\mathbf{k}}^{+,-}\chi_{\mathbf{k}}^{+,-}-x\delta V_{0}\chi_{\mathbf{k}}^{+,+}}{\omega_{\mathbf{k}}^{2}}\left[\frac{1}{\omega_{\mathbf{k}}}\tanh\left(\frac{\beta\omega_{\mathbf{k}}}{2}\right)-\frac{\beta}{2}\textrm{sech}^{2}\left(\frac{\beta\omega_{\mathbf{k}}}{2}\right)\right]\nonumber 
\end{eqnarray}
It is clear from Eq.\ref{eq:Average-n} that the self-consistency
equation for the number of particles is fulfilled when $\mu=V_{0}^{+,+}+V_{0}^{-,+}$,
which confirms our choice for the chemical potential. In addition,
the asymmetry factor is obtained from the next self-consistency equation:
\begin{eqnarray}
x & = & \frac{1}{N}\sum_{\mathbf{k}}\frac{2\chi_{\mathbf{k}}^{+,+}-x\delta V_{0}}{\tilde{\omega}_{\mathbf{k}}}\tanh\left(\frac{\beta\tilde{\omega}_{\mathbf{k}}}{2}\right)\label{eq:AssymetryFactor}\\
 &  & +2x\delta V_{0}\frac{1}{N}\sum_{\mathbf{k}}\frac{x\delta V_{0}\chi_{\mathbf{k}}^{+,+}-J_{\mathbf{k}}^{+,-}\chi_{\mathbf{k}}^{+,-}-J_{\mathbf{k}}^{-,+}\chi_{\mathbf{k}}^{-,+}}{\tilde{\omega}_{\mathbf{k}}^{2}}\left[\frac{\beta}{2}\textrm{sech}^{2}\left(\frac{\beta\tilde{\omega}_{\mathbf{k}}}{2}\right)-\frac{1}{\tilde{\omega}_{\mathbf{k}}}\tanh\left(\frac{\beta\tilde{\omega}_{\mathbf{k}}}{2}\right)\right]\nonumber 
\end{eqnarray}
\end{widetext}The solution to Eq.\ref{eq:AssymetryFactor} provides
the phase diagram. In Fig.\ref{fig:Full-Phase-diagram} we show this
phase diagram for the case of a dimerized chain. The comparison between
the uncorrelated and correlated phase boundaries show an interesting
feature: For the case $\lambda=t_{1}^{\prime}/t_{1}<1$ the homogeneous
charge phase is stabilized by correlations, however for the case $\lambda>1$
the CDW phase is the one stabilized. Furthermore, the case $\lambda<1$
shows two inequivalent solutions $x\neq0$ with different particle
distributions, in contrast with the case $\lambda>1$ with only one.
As we will discuss in the next section, each choice of $\lambda$
corresponds to a different topological phase.

\begin{figure}
\includegraphics[scale=0.6]{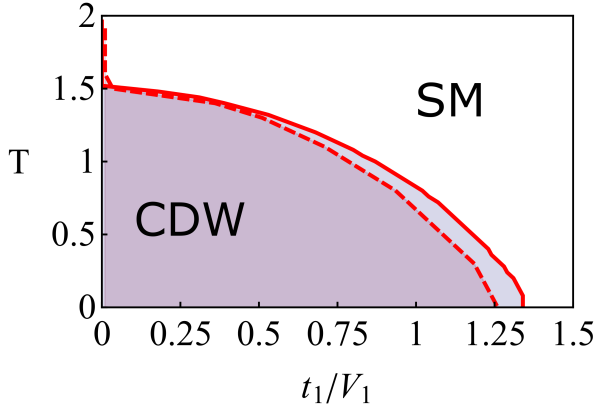}

\caption{\label{fig:HC-Full-PD}Phase diagram for the honeycomb lattice ($V_{2}=0$).
The continuous line corresponds to the phase boundary between the
CDW and the semi-metallic phase, when correlations are neglected.
The dashed line corresponds to the case including correlations. The
shift in the critical point towards a smaller value of $t_{1}/V_{1}$
means that the semi-metallic phase is stabilized by interactions.}
\end{figure}

In Fig.\ref{fig:HC-Full-PD} we compare the phase diagram for the
honeycomb lattice with and without correlations. It shows that the
semi-metallic phase becomes stabilized by 2-point correlations, shifting
the critical point to larger interactions.

It is interesting to notice that the effect of correlations decreases
as the dimensionality of the system increases, i.e., the correction
to the uncorrelated solution is more pronounced for the case of a
dimerized chain than for the honeycomb lattice. This is in agreement
with what one would expect from the critical behavior of low dimensional
systems, where quantum corrections should become more important as
the system decreases its dimensionality.

\section{Topological properties}

One of the most interesting aspects of a non-perturbative approach
is to ask whether some effects, which are genuinely non-perturbative,
can be captured. Here we study topological phases which lie out of
the non-interacting classification using the hierarchy of correlations,
and demonstrate that this approach provides an efficient method to
shed light on these elusive states of matter.

A fundamental object in the characterization of the topological properties
is the 2-point Green's function $G_{\mathbf{k}}^{\alpha,\beta}\left(t,t^{\prime}\right)$,
which classifies mappings from the Brillouin zone to the space of
$n$-dimensional matrices\cite{Gurarie1}. This is a very natural
way of classifying interacting topological phases, as in contrast
with the use of the Hamiltonian, the Green's function contains information
about symmetry broken phases and collective excitations. The Green's
function has been used for the characterization of edge states as
well\cite{GFunctions-BulkEdge}, and provides a new point of view
on how interactions can affect the topological properties. One of
the main consequences of this approach has been to show that interacting
systems can exhibit zeros in the Green's function, and they play a
similar role to poles (contributing to the topological index)\cite{GFTop1,GFTop2,No-Luttinger}.
Therefore, the interacting topological phase turns out to be defined
in terms of the poles and the zeros.

In this section we consider the full expression for the Green's function
when 2-point correlations are included (Eq.\ref{eq:TotalGF}), and
analyze the topological properties of the dimerized chain. We consider
a dimerized chain due to its simplicity, and to the advantage that
it has been thoroughly analyzed in previous works. It must be mentioned
that the expression used in this manuscript for the winding number
$\nu_{1}$ could breakdown if the Green's function has branch cuts
in the complex plane, instead of poles. Nevertheless, the models discussed
in this work do not seem to exhibit these properties and our description
should be correct.

In absence of interactions, the dimerized chain is characterized by
a 1D winding number classifying maps from the Brillouin zone ($\mathcal{S}^{1}$)
to the set of Hamiltonians with chiral symmetry ($\mathcal{S}^{1}$):
$k\rightarrow\hat{H}_{k}$. If we consider just nearest neighbors,
the winding number only depends on the ratio between the inter-dimer
and the intra-dimer hopping $\lambda=t_{1}^{\prime}/t_{1}$. When
$\lambda>1$, the winding number $\nu_{1}=1$ and the system is in
a topological phase that shows localized edge states for open boundary
conditions. On the other hand, when $\lambda<1$ the winding number
vanishes and the system does not exhibit edge states. Importantly,
as we have considered nearest neighbors only, the winding number cannot
be larger than $\nu_{1}=1$.

When interactions are included, the effects of correlations are encoded
in the Green's function. As previous works have shown, the 1D winding
number can be written in terms of $G_{k}^{\alpha,\beta}$, and in
absence of interactions, one recovers the well known expression in
terms of the Hamiltonian. The maps characterized by the winding number
correspond to $k\rightarrow\hat{G}_{k}$, where $\hat{G}_{k}$ is
the matrix with entries $G_{k}^{\alpha,\beta}$. Importantly, in the
presence of interactions we will show that $\hat{G}_{k}$ can lack
of chiral symmetry, as there are contributions to the diagonal elements
which are proportional to the asymmetry factor $x$ when sub-lattice
symmetry is spontaneously broken. This feature contrasts with the
fact that the Hamiltonian of the system contains chiral symmetry (given
as the product of time reversal and particle hole symmetry). This
means that the Green's function is clearly more appropriate for the
topological characterization than the Hamiltonian, as it contains
more information about the system. Concretely, when the system is
in the charge homogeneous phase ($x=0$) of Fig.\ref{fig:Full-Phase-diagram},
the Green's functions have chiral symmetry and the topological invariant
must be given by the winding number $\nu_{1}$. In contrast, for the
CDW phase ($x\neq0$) we have to classify mappings from the circle
to the sphere: $\mathcal{S}^{1}\rightarrow\mathcal{S}^{2}$, and as
they are always trivial, we always remain in a non-topological phase.
In conclusion, for the dimerized chain we only need to pay attention
to homogeneous charge phases, as they are the ones which can have
a non-vanishing topological invariant.

Finally, let us mention that we can use $\hat{G}_{k}\left(\omega=0\right)$,
rather than the full Green's function $\hat{G}_{k}\left(\omega\right)$
for the calculation of the winding number, as they both contain the
same topological properties in equilibrium\cite{SimplifiedTI}. This
feature allows us to highly improve the numerical calculation.

\begin{figure}[h]
\includegraphics[scale=0.6]{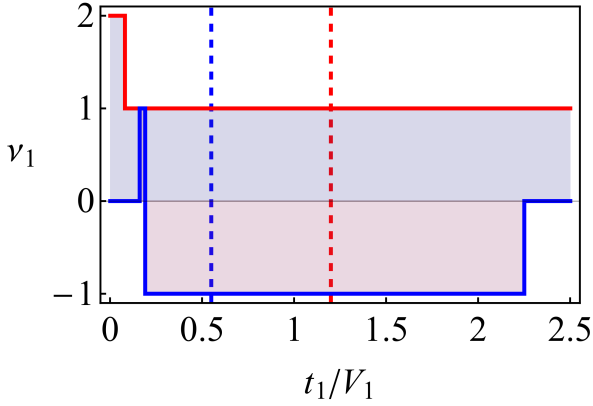}

\caption{\label{fig:Winding-number}Winding number $\nu_{1}$ of a dimerized
chain as a function of $t_{1}/V_{1}$ for different ratios of the
hopping $\lambda\in\left\{ 0.75,1.25\right\} $ (blue and red, respectively).
For large $t_{1}/V_{1}$ we recover the non-interacting winding number,
and as interactions increase, the winding number changes. The topological
phase (red) changes its winding number to $\nu_{1}=2$ when $t_{1}/V_{1}\simeq0.1$,
and the trivial phase (blue) changes to $\nu_{1}=-1$ at $t_{1}/V_{1}\simeq2.2$,
and to $\nu_{1}=0$ at $t_{1}/V_{1}\simeq1.15$. The vertical dashed
lines indicate the transition to the CDW phase from Fig.\ref{fig:Full-Phase-diagram}.
This calculation is performed assuming $T=0$.}

\end{figure}

The Green's functions at $\omega=0$ required for the characterization
are:
\begin{equation}
G_{k}^{\alpha,\beta}\left(0\right)=g_{k}^{\alpha,\beta}\left(0\right)-\frac{4J_{k}^{\alpha,\bar{\alpha}}\Lambda_{k}^{\bar{\alpha},\beta}\left(0\right)-2\alpha x\delta V_{0}\Lambda_{k}^{\alpha,\beta}\left(0\right)}{\tilde{\omega}_{k}^{2}}
\end{equation}
where $\Lambda_{k}^{\alpha,\beta}\left(\omega\right)\equiv\sum_{\sigma}g_{k}^{\sigma,\beta}\left(\omega\right)\chi_{k}^{\sigma,\alpha}$.
Furthermore, if we particularize to homogeneous charge phases ($x=0$),
they simplify to:
\begin{eqnarray}
G_{k}^{\alpha,\alpha}\left(0\right) & = & 0,\ G_{k}^{\alpha,\bar{\alpha}}\left(0\right)=\frac{J_{k}^{\alpha,\bar{\alpha}}}{\pi\tilde{\omega}_{k}^{2}}\left(1-\frac{\chi_{k}^{\alpha,\bar{\alpha}}}{J_{k}^{\bar{\alpha},\alpha}}\right)
\end{eqnarray}
and the matrix Green's function $\hat{G}_{k}$, whose entries are
the Green's function with different sub-lattice indices, has the following
general form:\begin{widetext}
\begin{equation}
\hat{G}_{k}\left(0\right)=\frac{1}{4\pi\left|J_{k}^{-,+}\right|^{2}}\left(\begin{array}{cc}
0 & J_{k}^{+,-}\left(1-\frac{\chi_{k}^{+,-}}{J_{k}^{-,+}}\right)\\
J_{k}^{-,+}\left(1-\frac{\chi_{k}^{-,+}}{J_{k}^{+,-}}\right) & 0
\end{array}\right)\label{eq:Chiral-GF}
\end{equation}
\end{widetext}where we know that the two response functions are related
according to $\chi_{k}^{-,+}=\left(\chi_{k}^{+,-}\right)^{\ast}$.
This matrix clearly has chiral symmetry, implemented by $\mathcal{C}=\sigma_{z}$.
Furthermore, we can see in Eq.\ref{eq:Chiral-GF} that the contributions
from the uncorrelated Green's function and the corrections due to
correlations are clearly separated. Then, any change with respect
to the winding number of the non-interacting system must come from
the response functions. Finally, we just need to apply the general
expression for the winding number in terms of the Green's function\cite{Volovik,SimplifiedTI}:
\begin{equation}
\nu_{1}=\frac{1}{4\pi i}\textrm{Tr}\oint\mathcal{C}\hat{G}_{k}^{-1}\partial_{k}\hat{G}_{k}dk
\end{equation}
When we make use of Eq.\ref{eq:Chiral-GF}, we find:\begin{widetext}
\begin{eqnarray}
\nu_{1} & = & \frac{1}{4\pi i}\oint\left(\frac{\partial_{k}J_{k}^{-,+}}{J_{k}^{-,+}-\chi_{k}^{+,-}}-\frac{\partial_{k}J_{k}^{+,-}}{J_{k}^{+,-}-\chi_{k}^{-,+}}\right)dk\label{eq:1DWinding}\\
 &  & +\frac{1}{4\pi i}\oint\left(\frac{2\chi_{k}^{-,+}\partial_{k}J_{k}^{+,-}-J_{k}^{+,-}\partial_{k}\chi_{k}^{-,+}}{J_{k}^{+,-}\left(J_{k}^{+,-}-\chi_{k}^{-,+}\right)}-\frac{2\chi_{k}^{+,-}\partial_{k}J_{k}^{-,+}-J_{k}^{-,+}\partial_{k}\chi_{k}^{+,-}}{J_{k}^{-,+}\left(J_{k}^{-,+}-\chi_{k}^{+,-}\right)}\right)dk\nonumber 
\end{eqnarray}
\end{widetext}This equation shows that the winding number in presence
of 2-point correlations has extra contributions encoded in the response
functions. Furthermore, the corrections are proportional to the convolutions
of the hybridization $\Delta_{k}^{\pm,\mp}$ with the interaction
potential $V_{k}^{\pm,\mp}$. Therefore, when the system is in the
homogeneous charge phase, the Green's function has chiral symmetry
and the topological invariant is generally given by Eq.\ref{eq:1DWinding}.
Furthermore, as this is a non-perturbative expression, it can be used
for arbitrary values of the interaction potential. We plot in Fig.\ref{fig:Winding-number}
the change in the topological index $\nu_{1}$, as a function of the
ratio between the hopping and the interaction strength $t_{1}/V_{1}$
.

It can be seen that the winding number changes with respect to the
non-interacting case as a function of the interaction strength, for
both the trivial ($t_{1}^{\prime}/t_{1}<1$) and the topological phase
($t_{1}^{\prime}/t_{1}>1$). More importantly, the winding number
can reach values larger than one, which is not possible for the dimerized
chain in absence of interactions. This is consistent with previous
works where the value of the winding number is expected to change
by unity\cite{Gurarie1}.

It is important to note that for the calculation of the winding number
we have assumed the homogeneous charge phase. Then one must compare
the phase diagram with the one in Fig.\ref{fig:Full-Phase-diagram}.
This comparison shows that we would not be able to detect the change
in the topological invariant when $t_{1}^{\prime}>t_{1}$ (the topological
phase in the non-interacting case), as the CDW phase would dominate.
However, it is interesting to see that the opposite would happen for
the trivial phase: In the absence of interactions and for $t_{1}^{\prime}<t_{1}$
the system is a normal insulator. Then, when interactions reach a
critical value $t_{1}/V_{1}\simeq2.2$, the insulator becomes topological
with winding number $\nu_{1}=-1$. As in the phase diagram of Fig.\ref{fig:Full-Phase-diagram},
correlations stabilize the homogeneous phase to $t_{1}/V_{1}\simeq0.55$,
the topological phase would exist for a wide range of the phase diagram.

We comment on one interesting aspect of the system behavior. The pole
structure of the Green's function has not been changed when correlations
are included, and one could think that the ``band structure'' of
the system did not close the gap during the topological transition
(this is a common feature of non-interacting topological insulators,
where the only way to change a topological index is by closing the
gap). Interestingly, in the presence of interactions the band structure
is not enough to characterize the topological properties. As has been
shown previously, the presence of interactions in the system introduces
zeros in the Green's function and they can compete with the poles.
These zeros are signatures of non-perturbative behavior\cite{1D-Top.Mott.Ins,Gurarie1,No-Luttinger}
and can annihilate the poles contribution, driving a topological phase
into a trivial one. More importantly, the zeros are not linked to
a gap closure, which means that we can have topological transitions
without closing the gap.

\begin{figure}
\includegraphics[scale=0.6]{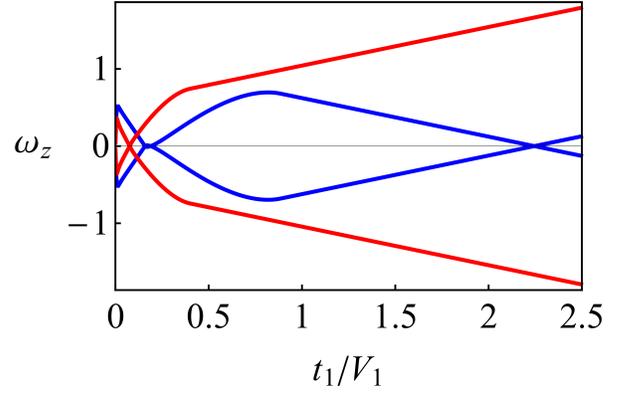}

\caption{\label{fig:Zeroes}Evolution of zeros of $\hat{G}_{k}\left(\omega\right)$
as a function of $t_{1}/V_{1}$. The blue(red) line corresponds to
the ratio $t_{1}/t_{1}^{\prime}=0.75\left(1.25\right)$. Note that
all changes in the topological invariant correspond to zeros of the
Green's function.}
\end{figure}

Here we prove that the change in the topological invariant $\nu_{1}$
is fully controlled by the appearance of zeros in $\hat{G}_{k}\left(\omega\right)$,
as has been discussed in previous works. To track their appearance
we calculate the determinant of $\hat{G}_{k}\left(\omega\right)$
for the case of the homogeneous charge phase:
\begin{equation}
\det\hat{G}_{k}\left(\omega\right)=\frac{\omega^{2}-4\left(J_{k}^{-,+}-\chi_{k}^{+,-}\right)\left(J_{k}^{+,-}-\chi_{k}^{-,+}\right)}{4\pi^{2}\left(\omega^{2}-4\left|J_{k}^{-,+}\right|^{2}\right)^{2}}
\end{equation}
From this expression one can see that zeros $\omega_{z}$ and poles
$\omega_{p}$ appear for:
\begin{eqnarray}
\omega_{z} & = & \pm2\sqrt{\left(J_{k}^{-,+}-\chi_{k}^{+,-}\right)\left(J_{k}^{+,-}-\chi_{k}^{-,+}\right)}\\
\omega_{p} & = & \pm2\left|J_{k}^{-,+}\right|
\end{eqnarray}
As the only quantity that depends on the interaction strength is $\omega_{z}$,
all topological changes should be linked to zeros of $\omega_{z}$.
In Fig.\ref{fig:Zeroes} we plot $\omega_{z}$ as a function of $t_{1}/V_{1}$,
and confirm that all changes in the winding number correspond to zeros
of the Green's function. 

\begin{figure}
\includegraphics[scale=0.5]{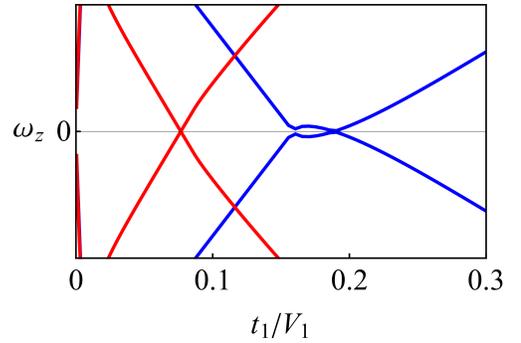}

\caption{\label{fig:Zoom-Zeroes}Zoom of Fig.\ref{fig:Zeroes} to small $t_{1}/V_{1}$.
It shows that even the fast change in the winding number from $\nu_{1}=-1$
to $\nu_{1}=1$ is captured by the zeros of the Green's function.}
\end{figure}

Furthermore, even the small region around $t_{1}/V_{1}\sim0.17$ in
Fig.\ref{fig:Winding-number}, where the winding number changes from
$\nu_{1}=-1$ to $\nu_{1}=1$ for the case $t_{1}/t_{1}^{\prime}=1.25$,
is reflected in the behavior of the zeros (see Fig.\ref{fig:Zoom-Zeroes}).

These results show that the topological characterization of strongly
correlated systems can be very different from the characterization
of non-interacting ones. This is due to the effect of correlations,
which can make the quasi-particle picture to be insufficient due to
the presence of collective excitations (now described by correlated
excitations). Nevertheless, the use of Green's functions and of non-perturbative
methods can be used to derive general results and capture the mechanisms
driving topological transitions. Furthermore, the calculation of the
motion of the zeros of the Green's function can be used as a complementary
``effective band structure'' to keep track of topological changes.

It must be mentioned that the presence of the $\nu_{1}=-1$ phase
for the dimerized chain at intermediate coupling ($t_{1}\sim V_{1}$)
is surprising, as it has not been predicted previously. Its appearance
corresponds to a regime away from both strong and weak coupling, and
one must be careful to confirm that higher corrections due to correlations
do not destroy this phase. In a future work we will address the numerical
detection of this phase using alternative methods.

\section{Conclusions}

In this work we have shown that a hierarchy of correlations can be
a useful tool in the study of many-body systems. We have shown its
connection with the $1/\mathcal{Z}$ expansion, and generalized it
to $\mathbf{k}$ space correlations, being the basic ingredient the
entanglement monogamy for strongly correlated systems. Although this
method is very general, here we focused on interacting spinless particles
in a dimerized chain, and in a honeycomb lattice. In general $\mathbf{k}$
space correlations scale differently from real space correlations,
and we have proved that the former can be more efficient, concretely
in systems with low coordination number. The results show that, for
the honeycomb lattice, correlations tend to stabilize the semi-metallic
phase, while correlations in the dimerized chain stabilize different
phases depending on the ratio between the intra-dimer and the inter-dimer
hopping ($t_{1}^{\prime}/t_{1}$). Finally, we have studied how this
approach can be applied to the study of topological phases with interactions.
One of the advantages is the simplicity of the expressions when 2-point
correlations are included, which provide information about the mechanisms
that can cause a change in the topological index when interactions
are present. Furthermore, we have shown that the Green's functions
in presence of correlations can have zeros, and that they are responsible
for these topological changes.

Future extensions of this work will address the role of higher order
correlations, the presence of long range interactions and non-equilibrium
systems. Furthermore, it would be interesting to understand the relation
between this approach and the renormalization group methods.

We would like to acknowledge F. Queisser, A. G. Grushin and P.C.E.
Stamp for useful comments and the critical reading of the manuscript.
This work was supported by NSERC, CIFAR, and PITP.

\bibliographystyle{phaip}
\bibliography{Hierarchy-QPT}

\begin{thebibliography}{10}

\bibitem{Bosonization}
A.~O. Gogolin, A.~Nersesyan, and A.~M. Tsvelik,
\newblock Cambridge University Press  (1998).

\bibitem{DynamicalMFT1}
H.~Aoki et~al.,
\newblock Rev. Mod. Phys. {\bf 86}, 779 (2014).

\bibitem{DynamicalMFT2}
A.~Georges, G.~Kotliar, W.~Krauth, and M.~J. Rozenberg,
\newblock Rev. Mod. Phys. {\bf 68}, 13 (1996).

\bibitem{QPT-Sachdev}
S.~Sachdev,
\newblock {\em Quantum Phase transitions},
\newblock Cambridge University Press, 2011.

\bibitem{PhaseTransitions-Landau}
V.~Ginzburg and L.~Landau,
\newblock Zh. Eksp. Teor. Fiz. {\bf 20}, 1064 (1950).

\bibitem{Renormalization-Wilson}
K.~G. Wilson,
\newblock Rev. Mod. Phys. {\bf 55}, 583 (1983).

\bibitem{QPT-Bernevig}
B.~A. Bernevig, T.~L. Hughes, and S.-C. Zhang,
\newblock Science {\bf 314}, 1757 (2006).

\bibitem{RevModPhys-TPT}
M.~Z. Hasan and C.~L. Kane,
\newblock Rev. Mod. Phys. {\bf 82}, 3045 (2010).

\bibitem{Ludwig-Classification}
A.~P. Schnyder, S.~Ryu, A.~Furusaki, and A.~W.~W. Ludwig,
\newblock Phys. Rev. B {\bf 78}, 195125 (2008).

\bibitem{Interac-TI-Class}
C.~Wang, A.~C. Potter, and T.~Senthil,
\newblock Science {\bf 343}, 629 (2014).

\bibitem{Top-Mott-Ins}
S.~Raghu, X.-L. Qi, C.~Honerkamp, and S.-C. Zhang,
\newblock Phys. Rev. Lett. {\bf 100}, 156401 (2008).

\bibitem{Top-Kondo-Ins}
M.~Dzero, K.~Sun, V.~Galitski, and P.~Coleman,
\newblock Phys. Rev. Lett. {\bf 104}, 106408 (2010).

\bibitem{Frac-QHI1}
T.~Neupert, L.~Santos, C.~Chamon, and C.~Mudry,
\newblock Phys. Rev. Lett. {\bf 106}, 236804 (2011).

\bibitem{Frac-QHI2}
N.~Regnault and B.~A. Bernevig,
\newblock Phys. Rev. X {\bf 1}, 021014 (2011).

\bibitem{Floquet-FCI}
A.~G. Grushin, A.~G\'omez-Le\'on, and T.~Neupert,
\newblock Phys. Rev. Lett. {\bf 112}, 156801 (2014).

\bibitem{Xin-Li2015}
J.-X.~L. Zhao-Long~Gu, Kai~Li,
\newblock Arxiv: {\bf 1512.05118} (2015).

\bibitem{TransverseIsing}
A.~Dutta et~al.,
\newblock {\em Quantum phase transitions in transverse field spin models: From
  statistical physics to quantum information},
\newblock Cambridge University Press, 2015.

\bibitem{Z-Exp1}
R.~S. Fishman and S.~H. Liu,
\newblock Phys. Rev. B {\bf 40}, 11028 (1989).

\bibitem{Z-Exp2}
H.~M. R\o{}nnow et~al.,
\newblock Phys. Rev. B {\bf 75}, 054426 (2007).

\bibitem{Queisser}
F.~Queisser, K.~V. Krutitsky, P.~Navez, and R.~Sch\"utzhold,
\newblock Phys. Rev. A {\bf 89}, 033616 (2014).

\bibitem{PhaseDiag-HC3}
C.~Weeks and M.~Franz,
\newblock Phys. Rev. B {\bf 81}, 085105 (2010).

\bibitem{PhaseDiag-HC2}
E.~V. Castro et~al.,
\newblock Phys. Rev. Lett. {\bf 107}, 106402 (2011).

\bibitem{PhaseDiag-HC1}
D.~D. Scherer, M.~M. Scherer, and C.~Honerkamp,
\newblock Phys. Rev. B {\bf 92}, 155137 (2015).

\bibitem{ExactDiag-HC}
S.~Capponi and A.~M. L\"auchli,
\newblock Phys. Rev. B {\bf 92}, 085146 (2015).

\bibitem{HC-Interactions1}
J.~Motruk, A.~G. Grushin, F.~de~Juan, and F.~Pollmann,
\newblock Phys. Rev. B {\bf 92}, 085147 (2015).

\bibitem{Concurrence}
V.~Coffman, J.~Kundu, and W.~K. Wootters,
\newblock Phys. Rev. A {\bf 61}, 052306 (2000).

\bibitem{Zubarev}
D.~N. Zubarev,
\newblock Soviet Physics Uspekhi {\bf 3}, 320 (1960).

\bibitem{Scherer2015}
D.~D. Scherer, M.~M. Scherer, and C.~Honerkamp,
\newblock Phys. Rev. B , 1 (2015).

\bibitem{Gurarie1}
V.~Gurarie,
\newblock Phys. Rev. B {\bf 83}, 085426 (2011).

\bibitem{GFunctions-BulkEdge}
A.~M. Essin and V.~Gurarie,
\newblock Phys. Rev. B {\bf 84}, 125132 (2011).

\bibitem{GFTop1}
Y.-Z. You, Z.~Wang, J.~Oon, and C.~Xu,
\newblock Phys. Rev. B {\bf 90}, 060502 (2014).

\bibitem{GFTop2}
S.~R. Manmana, A.~M. Essin, R.~M. Noack, and V.~Gurarie,
\newblock Phys. Rev. B {\bf 86}, 205119 (2012).

\bibitem{No-Luttinger}
K.~B. Dave, P.~W. Phillips, and C.~L. Kane,
\newblock Phys. Rev. Lett. {\bf 110}, 090403 (2013).

\bibitem{SimplifiedTI}
Z.~Wang and S.-C. Zhang,
\newblock Phys. Rev. X {\bf 2}, 031008 (2012).

\bibitem{Volovik}
M.~Silaev and G.~Volovik,
\newblock J. Low Temp. Phys. {\bf 161}, 460 (2010).

\bibitem{1D-Top.Mott.Ins}
T.~Yoshida, R.~Peters, S.~Fujimoto, and N.~Kawakami,
\newblock Phys. Rev. Lett. {\bf 112}, 196404 (2014).

\bibitem{Entang-monogamy}
A.~Osterloh and R.~Sch\"utzhold,
\newblock Phys. Rev. B {\bf 91}, 125114 (2015).

\end{thebibliography}

\begin{widetext}

\appendix

\section{Correlated part of the 4-point function}

In this appendix we include details of the calculation of the correlated
part of the 4-point function in the real space hierarchy $\mathcal{G}_{\mathbf{x}\mathbf{x}\mathbf{y};\mathbf{y}}^{\sigma\sigma\alpha;\beta}$.

The general equation of motion for the 4-point function is given by
($\mathbf{z}\neq\mathbf{y}$):
\begin{eqnarray}
\left(\omega+\mu\right)G_{\mathbf{z}\mathbf{z}\mathbf{y},\mathbf{y}}^{\gamma_{1}\gamma_{2}\alpha,\beta} & = & \langle f_{\mathbf{z},\gamma_{1}}^{\dagger}f_{\mathbf{z},\gamma_{2}}\rangle\frac{\delta_{\alpha,\beta}}{2\pi}+2\left(V_{\mathbf{z},\mathbf{z}}^{\gamma_{1},\gamma_{2}}-V_{\mathbf{z},\mathbf{z}}^{\gamma_{2},\gamma_{2}}\right)G_{\mathbf{z}\mathbf{z}\mathbf{y},\mathbf{y}}^{\gamma_{1}\gamma_{2}\alpha,\beta}\nonumber \\
 &  & +2\sum_{\mathbf{x},\sigma}\left(J_{\mathbf{z},\mathbf{x}}^{\gamma_{1},\sigma}G_{\mathbf{x}\mathbf{z}\mathbf{y},\mathbf{y}}^{\sigma\gamma_{2}\alpha,\beta}-J_{\mathbf{z},\mathbf{x}}^{\gamma_{2},\sigma}G_{\mathbf{z}\mathbf{x}\mathbf{y},\mathbf{y}}^{\gamma_{1}\sigma\alpha,\beta}-J_{\mathbf{y},\mathbf{x}}^{\alpha,\sigma}G_{\mathbf{z}\mathbf{z}\mathbf{x},\mathbf{y}}^{\gamma_{1}\gamma_{2}\sigma,\beta}\right)\nonumber \\
 &  & -2\sum_{\mathbf{x},\sigma}\left(V_{\mathbf{z},\mathbf{x}}^{\gamma_{1},\sigma}-V_{\mathbf{z},\mathbf{x}}^{\gamma_{2},\sigma}-V_{\mathbf{y},\mathbf{x}}^{\alpha,\sigma}\right)G_{\mathbf{z}\mathbf{z}\mathbf{x}\mathbf{x}\mathbf{y},\mathbf{y}}^{\gamma_{1}\gamma_{2}\sigma\sigma\alpha,\beta}
\end{eqnarray}
note that the 4-point function depends on the 6-point function now.
This reminds of the BBGKY hierarchy in statistical mechanics, and
a comparison between the $\mathcal{Z}^{-1}$ hierarchy and the BBGKY
has been previously dicussed in \cite{Queisser}. The correlated part
can be obtained by removing the uncorrelated one:
\begin{equation}
\mathcal{G}_{\mathbf{z}\mathbf{z}\mathbf{y},\mathbf{y}}^{\gamma_{1}\gamma_{2}\alpha,\beta}=G_{\mathbf{z}\mathbf{z}\mathbf{y},\mathbf{y}}^{\gamma_{1}\gamma_{2}\alpha,\beta}-\langle n_{\mathbf{z},\sigma}\rangle G_{\mathbf{y},\mathbf{y}}^{\alpha,\beta}
\end{equation}
\begin{eqnarray}
\left(\omega+\mu\right)\mathcal{G}_{\mathbf{z}\mathbf{z}\mathbf{y},\mathbf{y}}^{\gamma_{1}\gamma_{2}\alpha,\beta} & = & 2\left(V_{\mathbf{z},\mathbf{z}}^{\gamma_{1},\gamma_{2}}-V_{\mathbf{z},\mathbf{z}}^{\gamma_{2},\gamma_{2}}\right)G_{\mathbf{z}\mathbf{z}\mathbf{y},\mathbf{y}}^{\gamma_{1}\gamma_{2}\alpha,\beta}\\
 &  & +2\sum_{\mathbf{x},\sigma}\left(J_{\mathbf{z},\mathbf{x}}^{\gamma_{1},\sigma}G_{\mathbf{x}\mathbf{z}\mathbf{y},\mathbf{y}}^{\sigma\gamma_{2}\alpha,\beta}-J_{\mathbf{z},\mathbf{x}}^{\gamma_{2},\sigma}G_{\mathbf{z}\mathbf{x}\mathbf{y},\mathbf{y}}^{\gamma_{1}\sigma\alpha,\beta}\right)-2\sum_{\mathbf{x},\sigma}J_{\mathbf{y},\mathbf{x}}^{\alpha,\sigma}\left(G_{\mathbf{z}\mathbf{z}\mathbf{x},\mathbf{y}}^{\gamma_{1}\gamma_{2}\sigma,\beta}-\langle f_{\mathbf{z},\gamma_{1}}^{\dagger}f_{\mathbf{z},\gamma_{2}}\rangle G_{\mathbf{x},\mathbf{y}}^{\sigma,\beta}\right)\nonumber \\
 &  & -2\sum_{\mathbf{x},\sigma}\left(V_{\mathbf{z},\mathbf{x}}^{\gamma_{1},\sigma}-V_{\mathbf{z},\mathbf{x}}^{\gamma_{2},\sigma}\right)G_{\mathbf{z}\mathbf{z}\mathbf{x}\mathbf{x}\mathbf{y},\mathbf{y}}^{\gamma_{1}\gamma_{2}\sigma\sigma\alpha,\beta}+2\sum_{\mathbf{x},\sigma}V_{\mathbf{y},\mathbf{x}}^{\alpha,\sigma}\left(G_{\mathbf{z}\mathbf{z}\mathbf{x}\mathbf{x}\mathbf{y},\mathbf{y}}^{\gamma_{1}\gamma_{2}\sigma\sigma\alpha,\beta}-\langle f_{\mathbf{z},\gamma_{1}}^{\dagger}f_{\mathbf{z},\gamma_{2}}\rangle G_{\mathbf{x}\mathbf{x}\mathbf{y},\mathbf{y}}^{\sigma\sigma\alpha,\beta}\right)\nonumber 
\end{eqnarray}
which must be supplemented with the conservation law $i\partial_{t}\langle f_{\mathbf{z},\gamma_{1}}^{\dagger}f_{\mathbf{z},\gamma_{2}}\rangle=0$.
Using the expansion in correlations and keeping terms to $1/\mathcal{Z}$
order, we find:
\begin{eqnarray}
\left(\omega-\omega_{0,\alpha}\right)\mathcal{G}_{\mathbf{z}\mathbf{z}\mathbf{y},\mathbf{y}}^{\sigma\sigma\alpha,\beta} & = & 2\bar{n}_{\sigma}\left(1-\bar{n}_{\sigma}\right)g_{\mathbf{y},\mathbf{y}}^{\alpha,\beta}V_{\mathbf{y},\mathbf{z}}^{\alpha,\sigma}+2g_{\mathbf{y},\mathbf{y}}^{\alpha,\beta}\sum_{\sigma^{\prime},\mathbf{x}\neq\mathbf{y},\mathbf{z}}V_{\mathbf{y},\mathbf{x}}^{\alpha,\sigma^{\prime}}\langle n_{\mathbf{z},\sigma}n_{\mathbf{x},\sigma^{\prime}}\rangle_{C}
\end{eqnarray}
Using the correlated part of the 4-point function to characterize
the density-density correlations, we find after a Fourier transformation:
\begin{eqnarray}
\langle n_{\mathbf{k},\sigma}n_{\mathbf{k}^{\prime},\sigma^{\prime}}\rangle_{C} & = & -N\delta\left(\mathbf{k}+\mathbf{k}^{\prime}\right)\bar{n}_{\sigma}\left(1-\bar{n}_{\sigma}\right)\frac{V_{\mathbf{k}^{\prime}}^{\sigma^{\prime},\sigma}-\frac{V_{\mathbf{k}^{\prime}}^{\sigma^{\prime},\bar{\sigma^{\prime}}}V_{\mathbf{k}^{\prime}}^{\bar{\sigma^{\prime}},\sigma}}{\frac{1}{\frac{\beta}{2}\textrm{sech}^{2}\left(\frac{\beta\omega_{0,\sigma^{\prime}}}{2}\right)}+V_{\mathbf{k}^{\prime}}^{\sigma^{\prime},\sigma^{\prime}}}}{\frac{1}{\frac{\beta}{2}\textrm{sech}^{2}\left(\frac{\beta\omega_{0,\sigma^{\prime}}}{2}\right)}+V_{\mathbf{k}^{\prime}}^{\alpha,\alpha}-\frac{\left|V_{\mathbf{k}^{\prime}}^{\sigma^{\prime},\bar{\sigma^{\prime}}}\right|^{2}}{\frac{1}{\frac{\beta}{2}\textrm{sech}^{2}\left(\frac{\beta\omega_{0,\sigma^{\prime}}}{2}\right)}+V_{\mathbf{k}^{\prime}}^{\sigma^{\prime},\sigma^{\prime}}}}\\
 & = & -N\delta\left(\mathbf{k}+\mathbf{k}^{\prime}\right)\bar{n}_{\sigma}\left(1-\bar{n}_{\sigma}\right)\lambda_{\mathbf{k}}^{\sigma,\sigma^{\prime}}
\end{eqnarray}
Then, the statistical averages including correlations can be easily
found:
\begin{eqnarray}
\langle n_{\mathbf{x},\alpha}\rangle^{\mathbf{k}} & = & N\delta\left(\mathbf{k}\right)\left[\frac{1}{2}-\frac{1}{N}\sum_{\mathbf{q}}\frac{\omega_{0,\alpha}}{2\tilde{\omega}_{\mathbf{k}}}\tanh\left(\frac{\beta\tilde{\omega}_{\mathbf{k}}}{2}\right)\right]\\
 &  & +\delta\left(\mathbf{k}\right)4\beta^{2}\textrm{csch}^{3}\left(\beta\omega_{0,\alpha}\right)\sinh^{4}\left(\frac{\beta\omega_{0,\alpha}}{2}\right)\sum_{\sigma,\mathbf{q}}\left|V_{\mathbf{q}}^{\alpha,\sigma}\right|^{2}\bar{n}_{\sigma}\left(1-\bar{n}_{\sigma}\right)\nonumber \\
 &  & +4\beta^{2}\textrm{csch}^{3}\left(\beta\omega_{0,\alpha}\right)\sinh^{4}\left(\frac{\beta\omega_{0,\alpha}}{2}\right)\frac{1}{N}\sum_{\mathbf{q},\sigma,\sigma^{\prime}}V_{\mathbf{q}}^{\alpha,\sigma}V_{\mathbf{k}-\mathbf{q}}^{\alpha,\sigma^{\prime}}\langle n_{\mathbf{q},\sigma}n_{\mathbf{k}-\mathbf{q},\sigma^{\prime}}\rangle_{C}\nonumber \\
 & = & N\delta\left(\mathbf{k}\right)\left[\frac{1}{2}-\frac{1}{N}\sum_{\mathbf{q}}\frac{\omega_{0,\alpha}}{2\tilde{\omega}_{\mathbf{k}}}\tanh\left(\frac{\beta\tilde{\omega}_{\mathbf{k}}}{2}\right)\right]\nonumber \\
 &  & +N\delta\left(\mathbf{k}\right)4\beta^{2}\textrm{csch}^{3}\left(\beta\omega_{0,\alpha}\right)\sinh^{4}\left(\frac{\beta\omega_{0,\alpha}}{2}\right)\sum_{\sigma}\bar{n}_{\sigma}\left(1-\bar{n}_{\sigma}\right)\frac{1}{N}\sum_{\mathbf{q}}\left|V_{\mathbf{q}}^{\alpha,\sigma}\right|^{2}\left(1-\lambda_{\mathbf{q}}^{\sigma,\sigma}\right)\nonumber \\
 &  & -N\delta\left(\mathbf{k}\right)4\beta^{2}\textrm{csch}^{3}\left(\beta\omega_{0,\alpha}\right)\sinh^{4}\left(\frac{\beta\omega_{0,\alpha}}{2}\right)\sum_{\sigma}\bar{n}_{\sigma}\left(1-\bar{n}_{\sigma}\right)\frac{1}{N}\sum_{\mathbf{q}}V_{\mathbf{q}}^{\alpha,\sigma}V_{\mathbf{q}}^{\bar{\sigma},\alpha}\lambda_{\mathbf{k}}^{\sigma,\bar{\sigma}}\nonumber 
\end{eqnarray}
Interestingly one can calculate the integrals over $\mathbf{q}$ in
the continuum limit for both, nearest and next nearest neighbors interaction
in the dimerized chain and in the honeycomb lattice. The result is
surprisingly simple and does not depend on the details of the lattice
or dimension (only on the coordination number $\mathcal{Z}$):
\begin{equation}
\langle n_{\mathbf{x},\alpha}\rangle^{\mathbf{k}}=N\delta\left(\mathbf{k}\right)\left[\frac{1}{2}-\frac{1}{N}\sum_{\mathbf{q}}\frac{\omega_{0,\alpha}}{2\tilde{\omega}_{\mathbf{q}}}\tanh\left(\frac{\beta\tilde{\omega}_{\mathbf{q}}}{2}\right)\right]-N\delta\left(\mathbf{k}\right)\bar{n}_{\alpha}\left(1-\bar{n}_{\alpha}\right)\textrm{sinh}\left(\beta\omega_{0,\alpha}\right)
\end{equation}

\end{widetext}
\end{document}